\documentclass[12pt]{article}
\usepackage{amsmath}
\usepackage{natbib}
\usepackage{graphicx}
\usepackage{float}
\usepackage{color}
\usepackage{caption}
\usepackage{subcaption}
\usepackage{array}
\usepackage{tikz}
\usepackage{url}
\newcommand{\blind}{0}

\begin{document}

	\def\spacingset#1{\renewcommand{\baselinestretch}%
		{#1}\small\normalsize} \spacingset{1}

	
	\if0\blind
	{
		\title{\bf 
			CAT: a conditional association test for microbiome data using a leave-out approach}
		\author{Yushu Shi,  \\
			{\small Department of Population Health Sciences, Weill Cornell Medicine}\\[5pt]
			Liangliang Zhang,\\
			{\small Department of Population and Quantitative Health Sciences,}\\  {\small Case Western Reserve University} \\[5pt]
			Kim-Anh Do,\\
			{\small Department of Biostatistics,}\\  {\small The University of Texas MD Anderson Cancer Center}\\[5pt]
			Robert R.\ Jenq,\\
			{\small Department of Genomic Medicine,}\\  {\small The University of Texas MD Anderson Cancer Center}\\[5pt]
			and \\
			Christine B.\ Peterson\\
			{\small Department of Biostatistics,}\\  {\small The University of Texas MD Anderson Cancer Center}
		}
		\maketitle
	} \fi
	
	\if1\blind
	{
		\bigskip
		\bigskip
		\bigskip
		\begin{center}
			{\LARGE\bf Title}
		\end{center}
		\medskip
	} \fi
	
	\bigskip
	\begin{abstract}
		\textbf{Motivation:} In microbiome analysis, researchers often seek to identify taxonomic features associated with an outcome of interest. However, microbiome features are intercorrelated and linked by
		phylogenetic relationships, making it challenging to assess the association between an individual feature and an outcome. 
		Researchers have developed global tests for the association of microbiome profiles with outcomes using beta diversity metrics. These methods are popular since microbiome-specific metrics offer robustness to extreme values and can incorporate information on the phylogenetic tree structure. 
		Despite the popularity of global association testing, most existing methods for follow-up testing of individual features only consider the marginal effect and do not 
		provide relevant information for the design of microbiome interventions.\\
		\textbf{Results:} This paper proposes a novel conditional association test, \textbf{CAT}, which can account for other features and phylogenetic relatedness when testing the association between a feature and an outcome.
		\textbf{CAT} adopts a leave-out method, measuring the importance of a feature in predicting the outcome by removing that feature from the data and quantifying how much the association with the outcome is weakened through the change in the coefficient of determination $R^2$. By leveraging global tests including PERMANOVA and MiRKAT-based methods, \textbf{CAT} allows association testing for continuous, binary, categorical, count, survival, and correlated outcomes.
		We demonstrate through simulation studies that \textbf{CAT} can 
		provide a direct quantification of feature importance that is distinct from that of marginal association tests. We illustrate \textbf{CAT} with applications to two real-world studies on the microbiome in melanoma patients: one examining the role of the microbiome in shaping immunotherapy response, and one investigating the association between the microbiome and survival outcomes. Our results illustrate the potential of \textbf{CAT} to inform the design of microbiome interventions aimed at improving clinical outcomes. \\
		\textbf{Availability:} Our method has been implemented in the R package \texttt{CAT}, which is publicly available at \url{https://github.com/YushuShi/CAT}.\\
		\textbf{Contact:} yus4011@med.cornell.edu 
	\end{abstract}
	
	\vfill
	
	\newpage
	\spacingset{1.45} 

\section{Introduction}

The development of next-generation sequencing techniques has enabled high-resolution profiling of the human microbiome. The challenges of analyzing microbiome data include its high dimensionality and the structural relatedness of the observed features. As a starting point in assessing the link between the microbiome and an outcome, researchers often test the global association of a phenotype with the microbiome as a whole.
 Global association tests addressing this question typically employ microbiome-specific metrics (also called beta diversity measures) that are more robust to extreme values than classical Euclidean distances. Popular choices include Bray-Curtis dissimilarity \citep{BrayCurt57}, which is designed for count data, and weighted and unweighted UniFrac \citep{LozuKnig05,LozuHama07}, which incorporate information on the location of features in a phylogenetic tree.

For microbiome datasets with clearly identified sample groups, a popular global association test is PERMANOVA, which utilizes permutation testing to obtain a $p$-value for the null hypothesis that there is no difference in the location of the centroids across groups
\citep{Ande01}. PERMANOVA has been widely used in microbial analysis, as it is simple to apply and only requires observations of the outcome variable and the pairwise distances or dissimilarities between samples.  Moreover, recent adaptations of PERMANOVA can handle nested designs and correlated outcomes \citep{vegan}. 

Another popular global association test for microbiome data is MiRKAT, which is based on kernel machine regression \citep{ZhaoChen15}. One advantage of this method is that it can incorporate multiple candidate distance metrics to maximize power for a particular data set. In addition to continuous and binary response variables,  MiRKAT  has been extended to handle survival outcomes \citep{PlanZhan17} and correlated or dependent samples  \citep{ZhanXue18, KohLi19}. 
These global association 
methods have been widely applied in microbiome studies. However, they cannot provide inference on specific taxa.

Thus, a practical question in following up on a significant global association test result is to identify specific microbiome features that drive the global testing result.
However, existing methods for testing the effect of individual taxa often ignore the presence of related features and focus only on the marginal effect of the individual feature. In particular, popular differential abundance methods such as ALDEx2 \citep{FernReid14}, DESeq2 \citep{LoveHube14}, 
and ANCOM-BC \citep{LinPedd20}
all adopt a marginal testing framework. 

An inherent problem in marginal testing of microbiome data is nested discoveries,
where hits are linked within a taxonomic or phylogenetic hierarchy.
Taxonomic trees reflect the traditional labeling and organization of microorganisms into groupings such as family or genus, 
while phylogenetic trees reflect evolutionary history, with branch points corresponding to events that gave rise to differences in the genomic sequences. Both types of trees play a key role in understanding microbiome data.
Taxonomic labels serve as a basis for interpretation since they are standardized across studies.
 Phylogenetic trees are useful in analysis, as they encode  rich information on sequence similarity, which drives phenotypic and functional similarity. 

The relatedness among features can make it challenging to pinpoint which taxa play a critical role in influencing outcomes.
For example, when a genus is found to be significant,  the corresponding higher taxonomic units to which it belongs, such as family and order, also tend to be significant. However, the precise taxonomic level most relevant to the outcome is difficult to establish. A conditional test can provide direct quantification of the importance of a specific feature in contributing information not captured by other features in the data set, addressing a question with potentially greater biological importance than a marginal test. 

Notably, the challenge of correlated predictors exists in many high-dimensional datasets, yet is particularly prominent in microbiome data. In other settings, researchers have proposed rigorous definitions of feature-outcome independence. Here, we adopt \cite{CandFan18}'s definition, where a feature is said to be ``null" if and only if the outcome is independent of it conditionally on all other variables. 

 In this paper, we present
a novel conditional test, \textbf{CAT}, that provides a natural next step to follow up on a significant global association test result. \textbf{CAT} achieves 
the goal of assessing the importance of individual taxa
 while accounting for phylogenetic structure and other features in the data set. The remainder of the paper is organized as follows: Section \ref{sec:approach} describes the proposed \textbf{CAT} method in detail. Section \ref{sec:sim} demonstrates its performance using simulated data, and  Section \ref{sec:application} illustrates the method through applications to real datasets with binary and survival outcomes. Section \ref{sec:discussion} concludes the paper with a discussion.

\section{Approach} \label{sec:approach}

\begin{figure}[H]
\vskip-.3cm
\centering
\includegraphics[width=\linewidth]{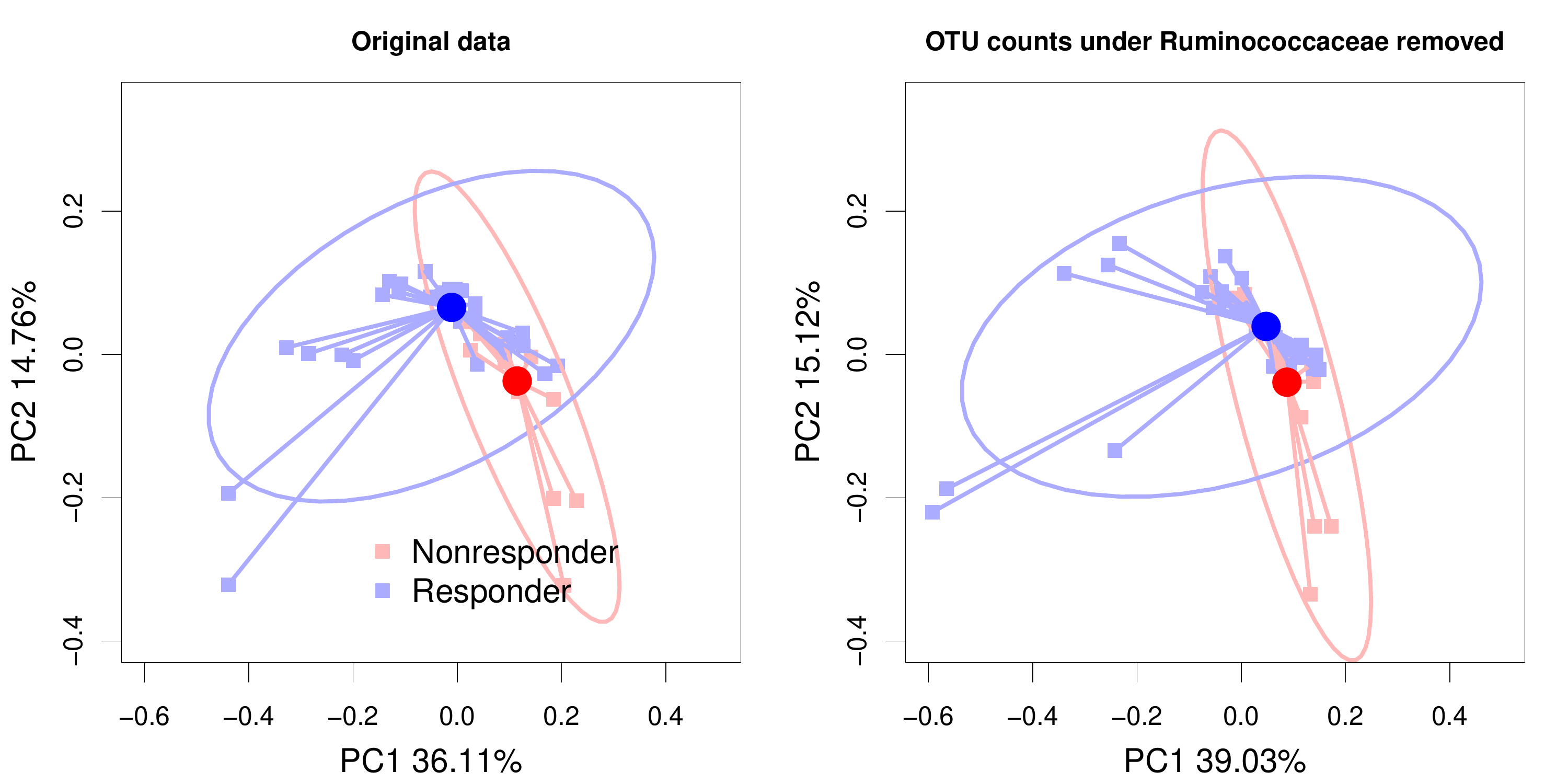}
	\caption{PCoA plots illustrating the global variation in the microbiome for melanoma patients who responded vs.\ did not respond to immunotherapy  
	 \citep{GopaSpen18}. We depict both the original data (left) and the modified data after removing counts belonging to the family Ruminococcaceae. The ellipses represent the 95\% confidence regions. }
	\label{DeepakPCoA}
\vskip-.3cm
\end{figure}

We begin with an illustration highlighting the motivation behind our approach. In the left panel of Figure \ref{DeepakPCoA}, we show a PCoA plot based on weighted UniFrac distance that depicts variation in microbiome composition between melanoma patients that responded to immunotherapy vs.\ those that did not.
The 95\% confidence regions for each group indicate there may be global differences in the microbiome profiles between the groups.
In the right panel, we artificially removed counts belonging to the family Ruminococcaceae; the two clusters of patients become less separated in the PCoA plot, with a reduced distance between centroids. Removing Ruminococcaceae from the data weakened the global association between the microbiome and response, suggesting that Ruminococcaceae may play an important role in driving the global association results. In the remainder of this section, we describe our proposal for a formal testing procedure aimed at quantitatively describing this phenomenon.

In our proposed approach, 
we start with the finest resolution features, corresponding to the leaf nodes in the taxonomic tree. For microbiome data derived from profiling of the 16S rRNA gene, these features are typically defined as Amplicon Sequence Variants (ASVs) or operational taxonomic units (OTUs). Our  method may be applied as well to features derived from whole metagenome sequencing (WGS).
By comparing the representative sequence for each feature against an established reference library,
the feature can be assigned a taxonomic classification. Taxonomic levels from broad to specific follow the sequence \textsl{kingdom}, \textsl{phylum}, \textsl{class}, \textsl{order}, \textsl{family}, \textsl{genus}, and \textsl{species}. 
Based on the taxonomic assignments, one can draw a taxonomic tree reflecting the relatedness of all the features in the data set. 


Most existing methods for identifying individual feature  associations from WGS or 16S data focus on marginal associations and do not quantify how much individual features contribute to the results from global association testing. To fill this gap, we propose \textbf{CAT}, which tests the association between specific features and outcomes while conditioning on the tree structure and the abundance of other features in the tree.  
The \textbf{CAT} method is rooted in the coefficient of determination $R^2$ for global microbiome association tests.
 \textbf{CAT} estimates the change in $R^2$ for a global test using the original dataset vs.\ a modified data set with the taxon of interest removed, and obtains a $p$-value
through a bootstrap procedure, which entails sampling from the data with replacement \citep{efron1994}.
One widely used global association test is the nonparametric PERMANOVA method \citep{Ande01, Ande17}. We begin by briefly reviewing this test, which serves as a starting point for our proposed method.


	Consider the $n \times n$ matrix of pairwise distances between $n$ observations $\mathbf{D}=[d_{ij}],$ where $d_{ij}$ 
represents the distance between observation $i$ and observation $j.$
We transform $\mathbf{D}$ to a new matrix $\mathbf{A}=[a_{ij}]=[-\frac{1}{2}d_{ij}^2]$, and center $\mathbf{A}$ to get Gower's centered matrix\\ \vskip-.6cm
\begin{equation*} 
\mathbf{G} = \Big(\mathbf{I}-\frac{\mathbf{11'}}{n}\Big)\mathbf{A}\Big(\mathbf{I}-\frac{\mathbf{11'}}{n}\Big),\end{equation*} 
\noindent where $\mathbf{I}$ represents the $n \times n$ identity matrix, and $\mathbf{11'}$ represents an $n \times n$ matrix of all 1s.
With an $n \times g$ design matrix $\mathbf{X}$ providing information on $g$ covariates, we can compute the hat matrix $\mathbf{H}=\mathbf{X}'(\mathbf{X}'\mathbf{X})^{-1}\mathbf{X}$. From the hat matrix, we can further calculate the total sum-of-squares ($SS_T$), the among-group sum-of-squares ($SS_A$), and the residual sum-of-squares ($SS_R$) as in MANOVA: 
$$SS_T=\mathrm{tr}(\mathbf{G}), \quad SS_A=\mathrm{tr}(\mathbf{HG}), \text{ and }SS_R=\mathrm{tr}\big[(\mathbf{I}-\mathbf{H})\mathbf{G}\big].$$ 
Just as in MANOVA, the coefficient of determination $R^2$ can be calculated as the ratio of the sum of squares between groups ($SS_A$) to the sum of squares total ($SS_T$). It provides an indication of the strength of the relationship between the outcome variable and the microbiome profiles, with a value closer to 1 indicating a stronger relationship.

	We now describe how to apply \textbf{CAT} to test the conditional association between the outcome of interest and a specific taxon. Let $X$ denote the outcome vector for $n$ observations. Let $\mathbf{Z}$ represent the $n \times m$ matrix with the observed counts for the finest-resolution microbiome features, which correspond to the leaf nodes in a taxonomy tree $\mathcal{T}$ with $m$ leaves. We denote the set of leaf nodes for the full tree $\mathcal{L}(\mathcal{T}) = \{1, \ldots, m\}$. For any internal node in the tree $t$, we let $\mathcal{L}(t)$ denote the leaf nodes corresponding to its descendants.
 Given these definitions, we lay out the steps of the \textbf{CAT} procedure as follows:
\begin{enumerate}
	\setlength\itemsep{0em}
	\item Calculate the $n \times n$ sample pairwise distance matrix $\mathbf{D}$ for the original data matrix $\mathbf{Z}$.
	\item Perform PERMANOVA using $\mathbf{D}$ and $X,$ and obtain a coefficient of determination $R^2$ for the outcome of interest.
	\item For the specific taxon $t$ being tested by \textbf{CAT}, generate a new data matrix $\mathbf{Z}^*$ by converting all the elements of $\mathcal{L}(t)$ to have $0$ counts.
	\item Calculate a new pairwise distance matrix $\mathbf{D}^*$ using the modified data matrix $\mathbf{Z}^*$.
	\item Perform PERMANOVA using $\mathbf{D}^*$ and get a new coefficient of determination $R^{2*}$ for the outcome of interest.
    \item Perform bootstrap sampling, selecting the matching sample from the pairwise distance matrix $\mathbf{D}$, the modified distance matrix $\mathbf{D}^*$, and the outcome $X$ for $B$ bootstrap samples. For each sample, compute the coefficients of determination for the original and modified distances, $R^2_{(1)},R^2_{(2)},\dots,R^2_{(B)}$ and $R^{2*}_{(1)}, R^{2*}_{(2)},\dots, R^{2*}_{(B)}.$
     \item Compute the differences between the original $R^2$ and the leave-taxon-out $R^{2*}$ for all $B$ bootstrap samples. The estimated $p$-value is the proportion of the $R^2$ differences that are less than zero: $$\hat{p}=\frac{1}{B}\sum_{i=1}^B I(R^{2}_{(i)}-R^{2*}_{(i)}<0).$$ 
\end{enumerate}

\noindent
Figure \ref{illustration} provides a toy example to illustrate how \textbf{CAT} converts the leaf counts under a specific taxon $t$ to $0$ in Step 3 of the procedure. 
Suppose that taxon $t$ is strongly associated with the outcome of interest  and that this association is not captured by other features in the tree. In that case, the removal of the counts descending from $t$ will decrease the $R^2$ in the PERMANOVA test. In contrast, removing a non-discriminating taxon would minimally affect the $R^2$ of the PERMANOVA test. 

\begin{figure}[H]
	
	\begin{subfigure}{\linewidth}
		\caption{ Original data}
		\begin{tikzpicture}
	\node at (0,0) {\includegraphics[width=0.475\textwidth]{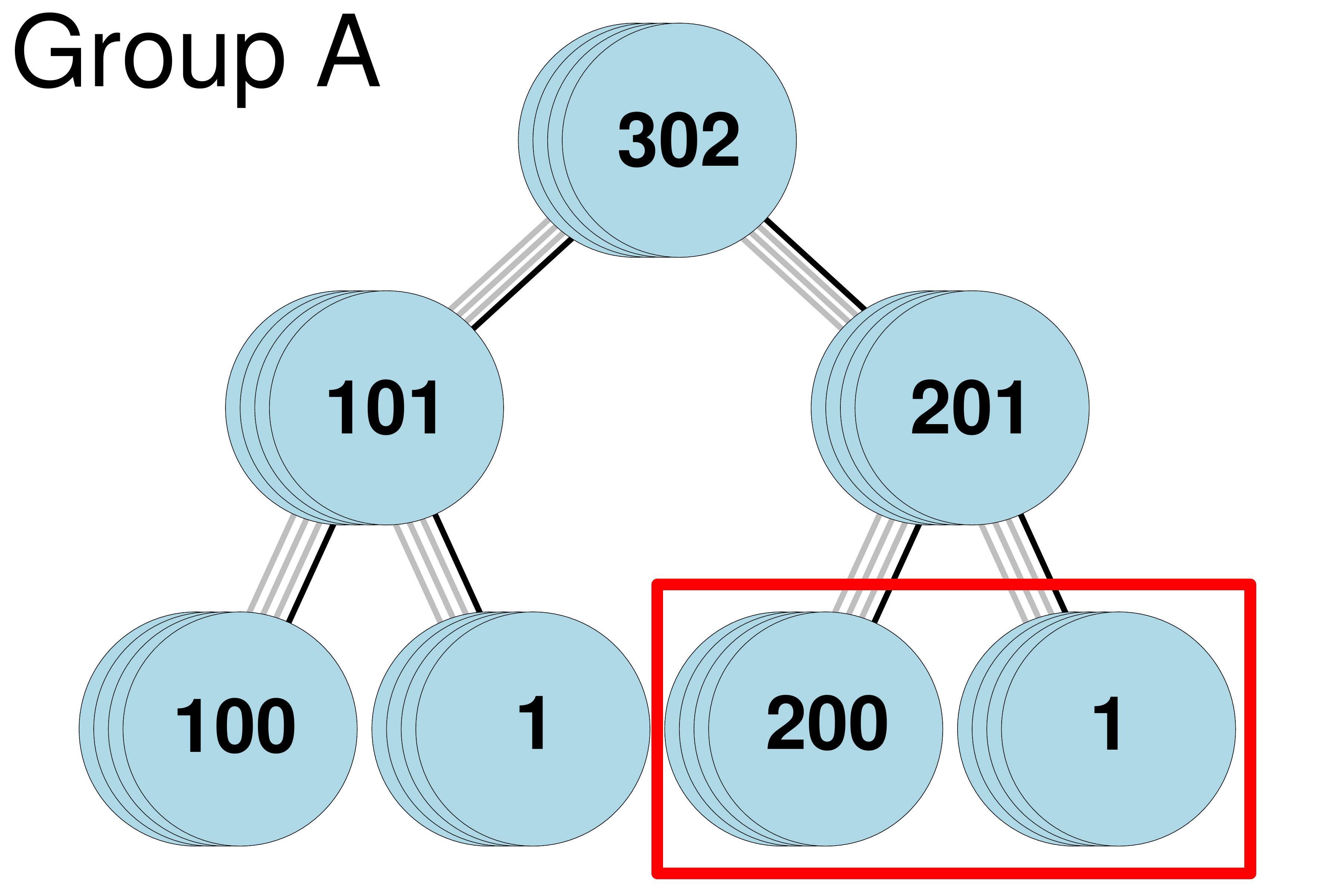}};
		\node at (1.2, .6) {{\color{red}\large{$\ast$}} \large{$t$}};
		\end{tikzpicture}
		\begin{tikzpicture}
	\node at (0,0) {\includegraphics[width=0.475\textwidth]{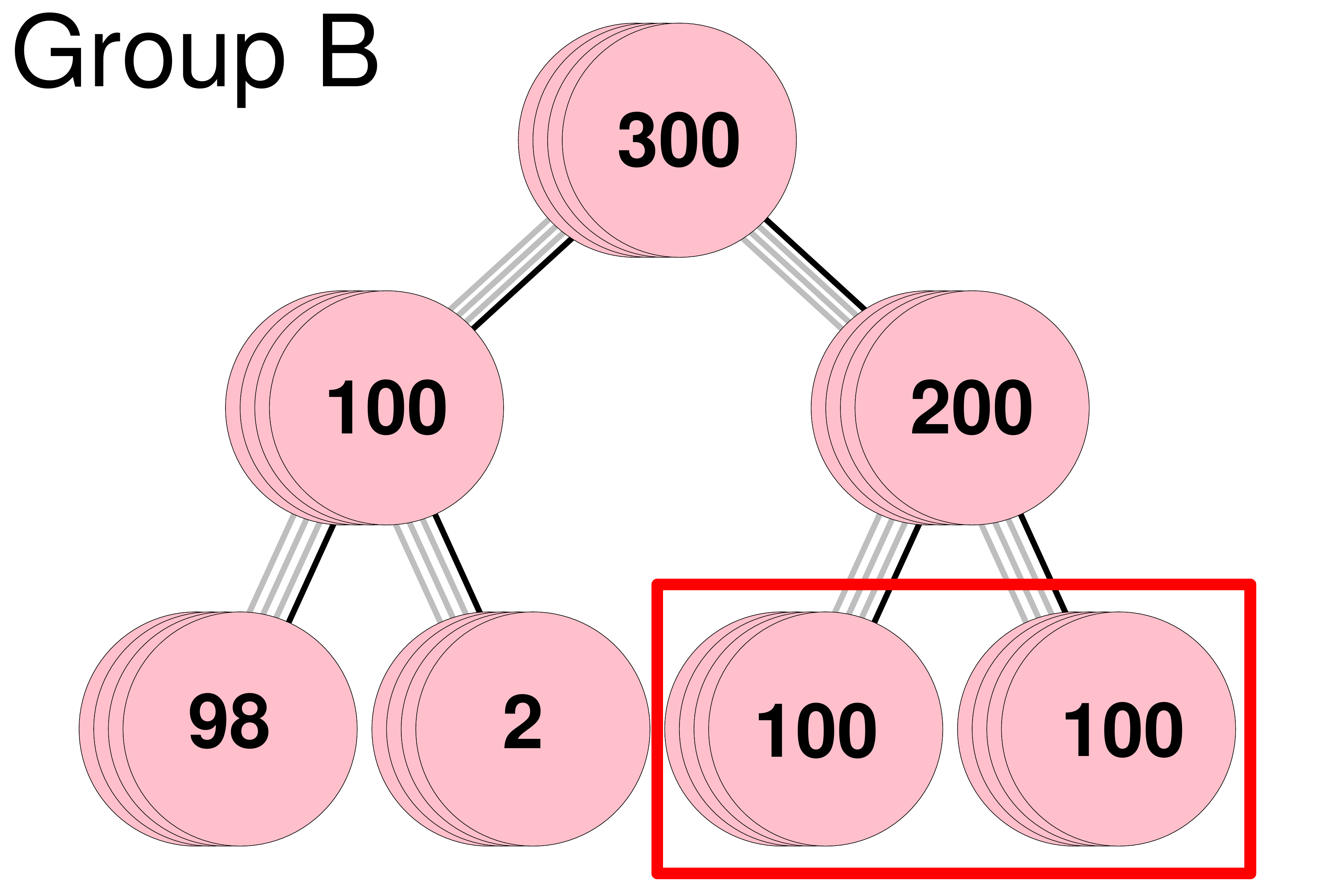}};
	\node at (1.2, .6) {{\color{red}\large{$\ast$}} \large{$t$}};
		\end{tikzpicture}
	\end{subfigure}
	
	\end{figure}
	
	\begin{figure}[H]\ContinuedFloat
	\begin{subfigure}{\linewidth}
		\caption{Modified data after converting the leaf nodes descending from taxon $t$ to $0$}
		\begin{tikzpicture}
	\node at (0,0) {\includegraphics[width=0.475\textwidth]{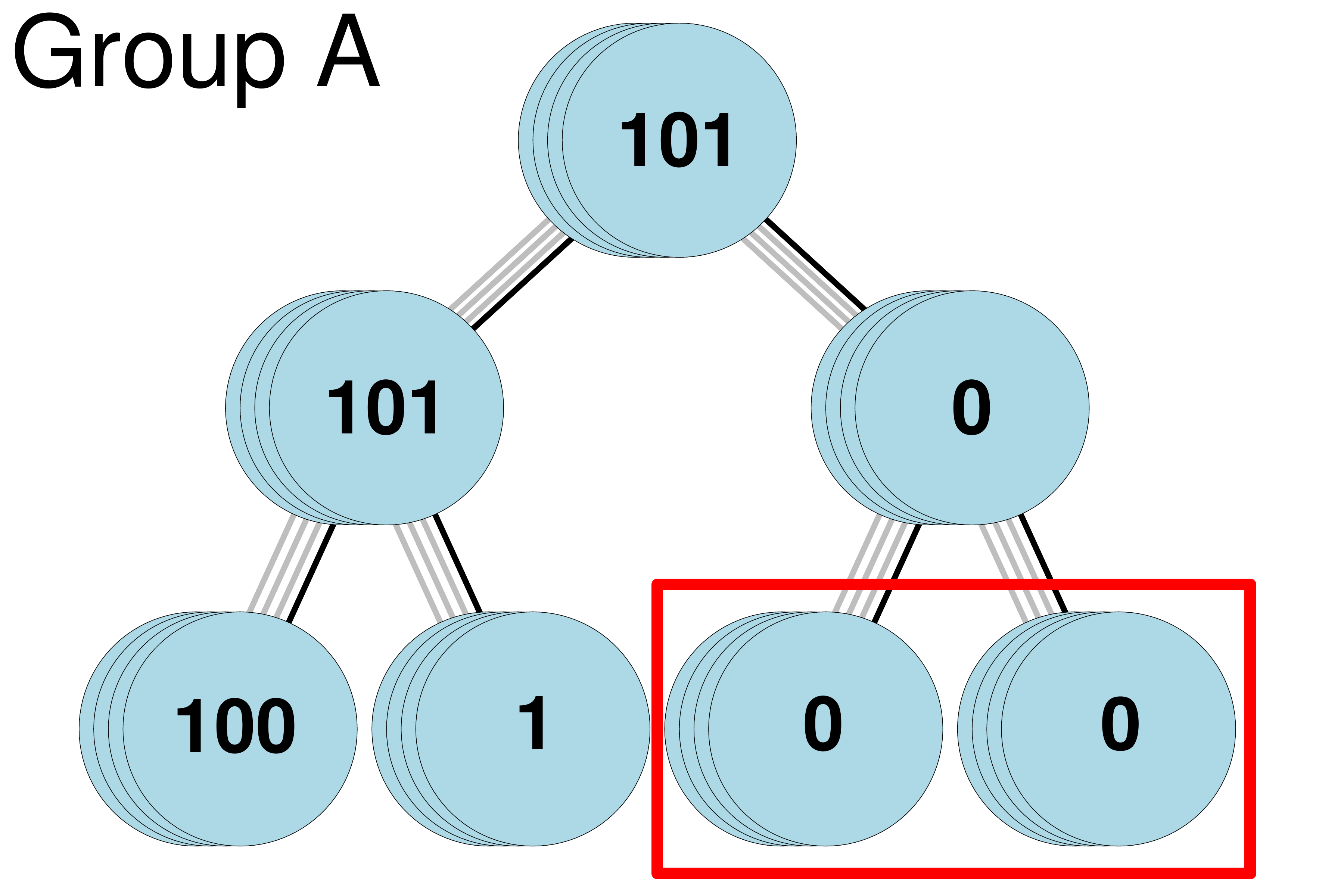}};
		\node at (1.2, .6) {{\color{red}\large{$\ast$}} \large{$t$}};
		\end{tikzpicture}
		\begin{tikzpicture}
	\node at (0,0) {\includegraphics[width=0.475\textwidth]{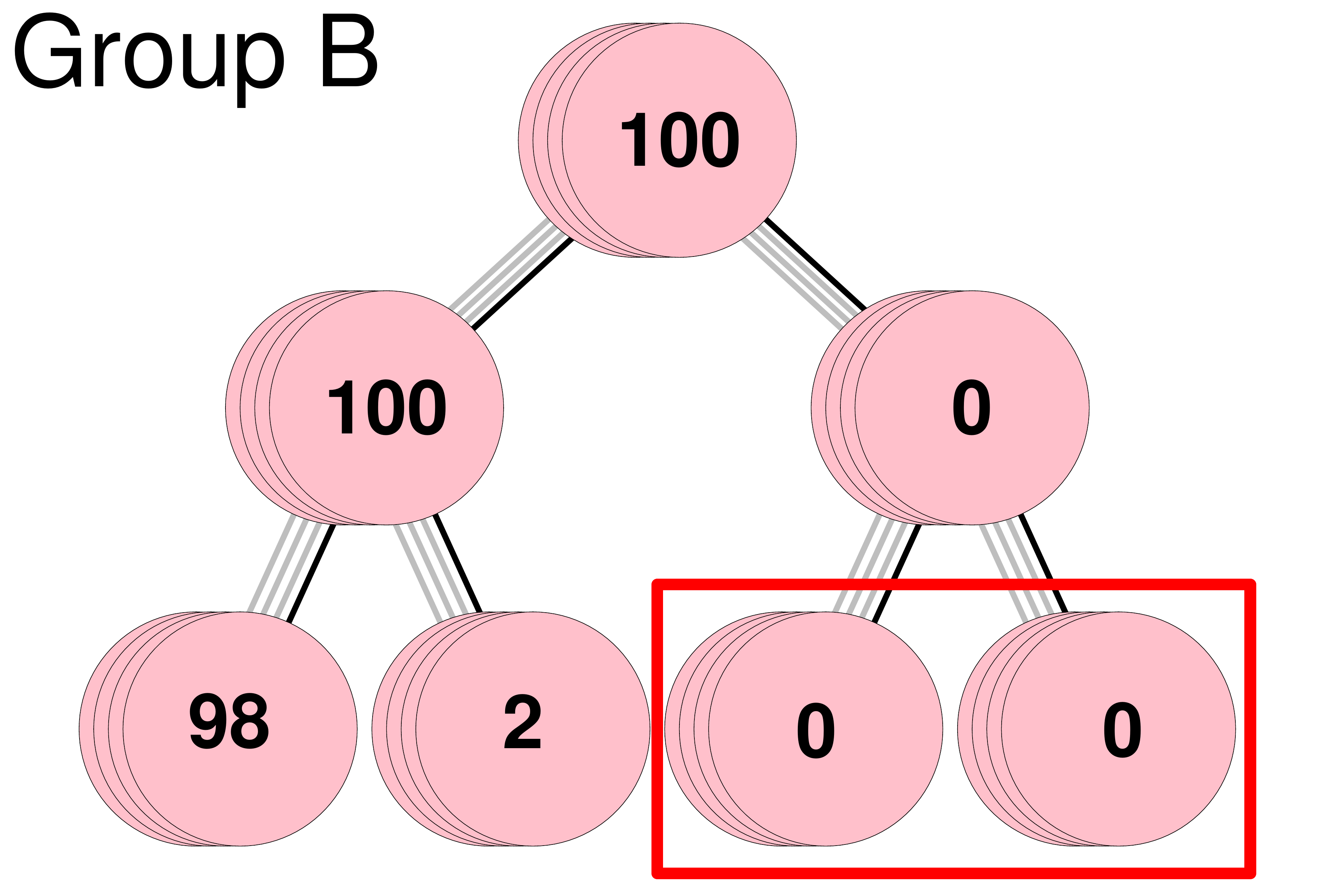}};
		\node at (1.2, .6) {{\color{red}\large{$\ast$}} \large{$t$}};
		\end{tikzpicture}
	\end{subfigure}
	
	\caption{\small A schematic plot showing how to convert the descending counts to $0$ for a particular taxon. Taxa in the red rectangles are children of the taxon of interest.}
	\label{illustration}
\end{figure}

\subsection{Conditional testing for MiRKAT}

The MiRKAT method \citep{ZhaoChen15}, which has been extended to handle survival outcomes \citep{PlanZhan17} and correlated or dependent samples \citep{ZhanXue18, KohLi19}, offers a powerful global association test based on the kernel regression framework.


For the simplest situation where the outcome of interest $Y$ is continuous, the model can be expressed as
$$
y_i=\beta_0+\boldsymbol{\beta}\mathbf{x}_i+f(\mathbf{z}_i)+\varepsilon_i, \quad i=1,2,\dots,n,
$$
where $y_i$ is the outcome for the $i$th subject, $\beta_0$ is the intercept term, $\boldsymbol{\beta}$ is a vector of regression coefficients, and
 $\mathbf{x}_i$ is a vector of covariates unrelated to the microbiome.
 The microbiome information for the $i$th sample is characterized by $\mathbf{z}_i$, and $f(\mathbf{z}_i)$ is the output from a reproducing kernel Hilbert space $\mathcal{H}_k$. The microbiome association test is equivalent to testing $f(\mathbf{z})=0$.
Given the pairwise distances, the kernel matrix between observations is taken as $
\mathbf{K} = -1/2\Big(\mathbf{I}-\frac{\mathbf{11'}}{n}\Big)\mathbf{A}\Big(\mathbf{I}-\frac{\mathbf{11'}}{n}\Big)$,
 with a correction when necessary to ensure that the matrix is positive semi-definite.

\cite{Zhan19} showed that the squared MiRKAT statistic is proportional to the $R^2$ statistic, or coefficient of determination, up to a constant factor. In this setting, $R^2$ characterizes the fraction of variability in outcome similarity explained by microbiome similarity.  This permits the use of \textbf{CAT} for testing conditional association for a particular taxon.  To implement this approach, the MiRKAT procedure can replace PERMANOVA in Steps 2, 5, and 6 of the \textbf{CAT} procedure. When multiple distance metrics are used, users can take the maximum of the $R^2$ from different metrics. More broadly, any valid global 
testing method can be used in these steps.

	\section{Simulation study} \label{sec:sim}
	In this section, we illustrate the utility of \textbf{CAT} as a follow-up to global testing and
 compare results from \textbf{CAT} with those from existing marginal testing approaches on simulated data. We develop realistic simulation scenarios by starting from a real microbiome data set, which will be examined more closely in Section \ref{sec:application}. We  adopt a "spike-in" method to control the taxa driving the cross-group differences. 
	
		\subsection{Data generation}
	To construct our simulation, we first obtained the 16S sequencing data from  \cite{GopaSpen18}, which examined the association between the gut microbiome and response to immunotherapy in melanoma patients. This is the same data set illustrated in Figure \ref{DeepakPCoA}. The original data included $43$ patients; $30$ responded to therapy, while the rest were non-responders.
 The sequencing depths per sample in this study 
 had a mean of 48,765. To quantify features from the raw sequencing data, we applied the UNOISE2 function \citep{Edga16} to the 16S rRNA gene sequences, identifying 1,455 ASVs. Given the  sequence for each ASV, we then applied the FastTree algorithm \citep{PricDeha09} to build a phylogenetic tree. 

	In our simulation set-up, we assume there are two groups (group 0 and group 1), each with 31 observations. We use the mean sequencing depth of the melanoma data as the number of sequences for each sample. The steps for generating the simulated data are:
	
	\begin{figure}[H]
 \centering
		\begin{subfigure}{\linewidth}
			\caption{Family Porphyromonadaceae and related taxa}
			\vskip-.16cm 
   \includegraphics*[width=0.46\textwidth,trim={0 0 0 1cm},clip]{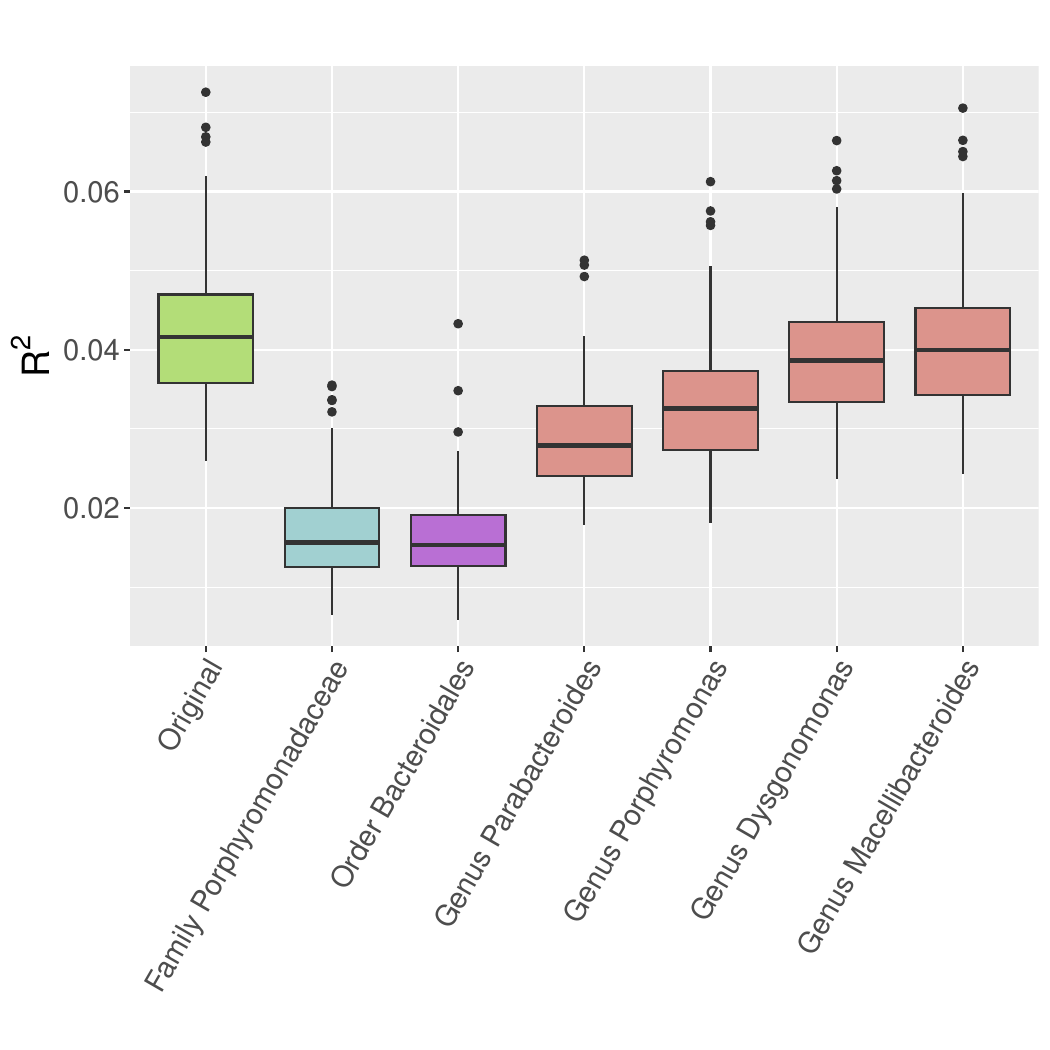}
			\includegraphics*[width=0.48\textwidth,trim={0 0 0 1cm},clip]{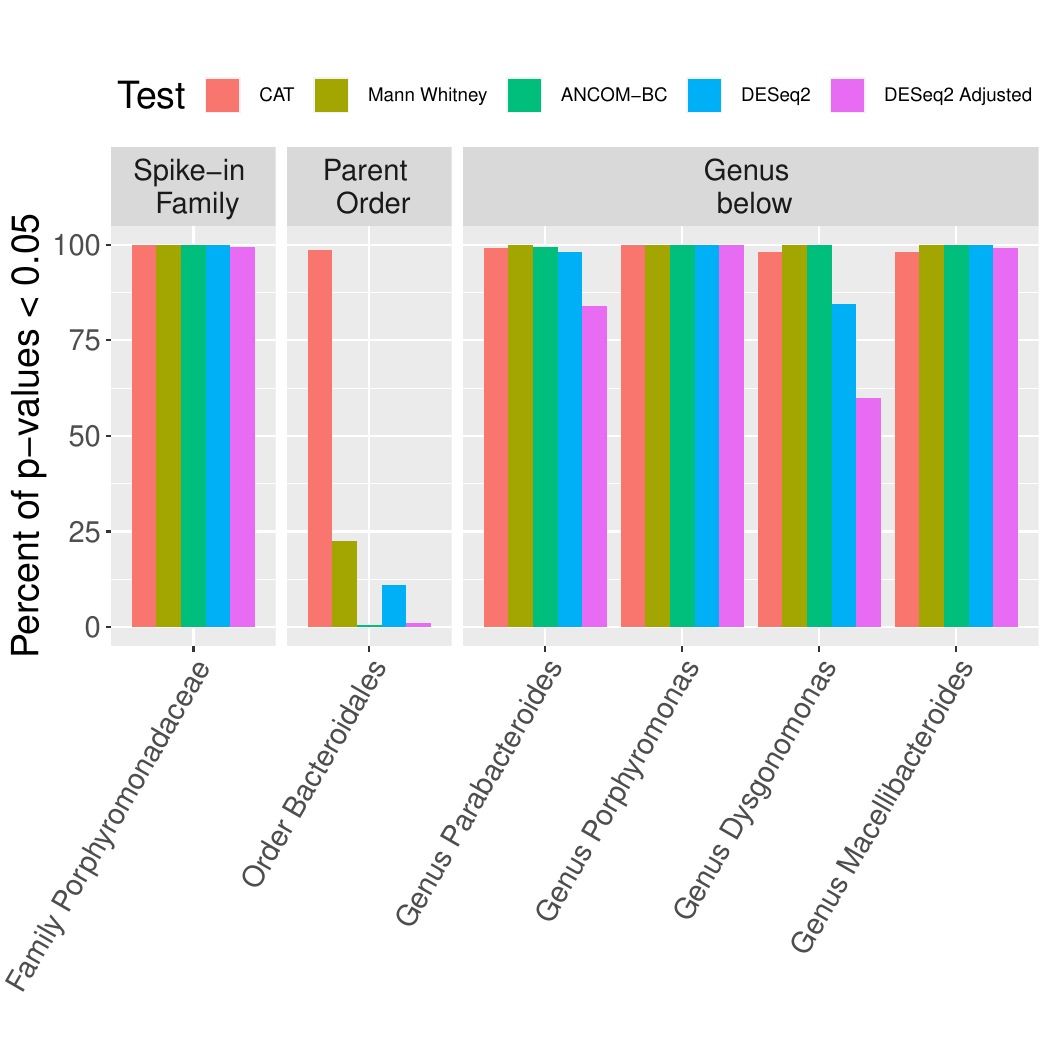}
		\end{subfigure}

		\begin{subfigure}{\linewidth}
			\vskip-.4cm  \caption{Family Lachnospiraceae and related taxa}
				\vskip-.16cm 
      \includegraphics*[width=0.46\textwidth,trim={0 0 0 1cm},clip]{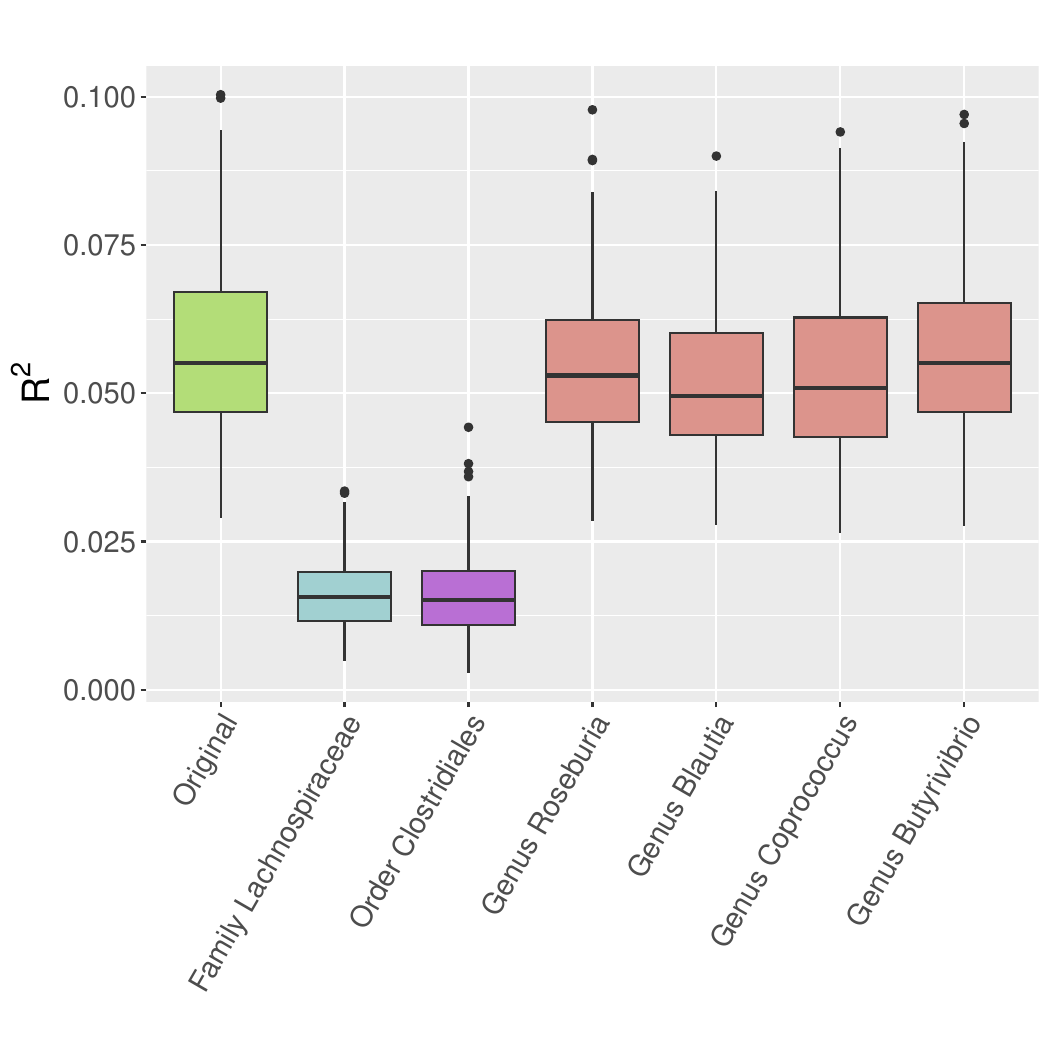}
			\includegraphics*[width=0.48\textwidth,trim={0 0 0 1cm},clip]{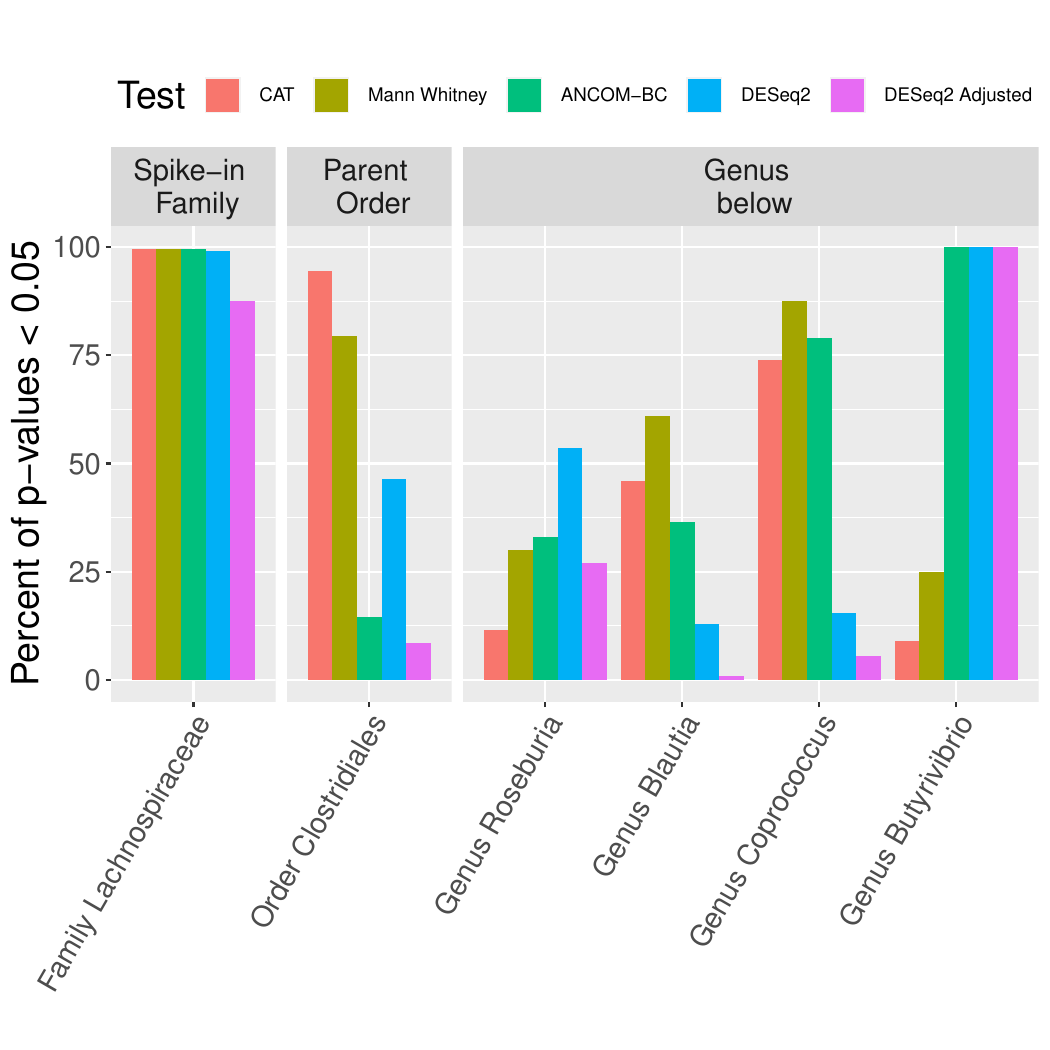}
		\end{subfigure}
		
		\caption{Boxplots of the $R^2$ values (left) and 
  barplots of the percentage of $p$-values less than 0.05 from \textbf{CAT}, Mann-Whitney, ANCOM-BC, and DeSeq2 (right) over 200 simulated datasets. The first panel in each subplot in the right column represents results from the manipulated family, followed by the order above and the child genera below.}
		\label{Sim}
	\end{figure}

\begin{enumerate}
	\item For each group, set the expected abundance of each microbiome feature to that of the marginal distribution of the melanoma dataset. 
	\item Generate the number of sequences for each ASV from a Dirichlet multinomial distribution with the sum of parameters for the Dirichlet distribution set to $62$. (In the melanoma dataset, if we assume the data are from two Dirichlet multinomial distributions, the sums of the parameters are estimated to be $70$ for the responder group and $54$ for the non-responder group.) 
	\item For group 1, add a random number generated from a Poisson distribution with parameter $\lambda$ to the number of sequences for the ASVs belonging to the feature being "spiked-in".
\end{enumerate}

\noindent Varying the parameter $\lambda$ affects the signal strength; we test the performance of our method with $\lambda$ set to 5, 10, 30, 50, and 70. The simulation study has two scenarios corresponding to three differential features: family Porphyromonadaceae, which accounts for 32 ASVs and 3.5\% of the sequences in the melanoma dataset; and family Lachnospiraceae (346 ASVs, 14.0\% of sequences). 
We chose moderately abundant families to best illustrate  differential performance across methods. The range of family abundances varies widely and is highly skewed, with Bacteroidaceae comprising the highest proportion ($41.98\%$) and Aeromonadaceae the lowest ($0.00005\%$), resulting in a median abundance of $0.01\%.$

	\subsection{Methods compared}
	In applying \textbf{CAT}, we set the number of bootstrap samples to 1000.
	To fully leverage the phylogenetic information, we use the weighted UniFrac distance, which accounts for ASV abundance, as well as the topology and branch lengths of the phylogenetic tree. The phylogenetic tree used for calculating the weighted UniFrac distances is the phylogenetic tree of \cite{GopaSpen18}'s dataset. 
 To illustrate the utility of \textbf{CAT} as a follow-up to global hypothesis testing, we show the distribution of $R^2$ values in the real and modified data sets as side-by-side boxplots. Although  \textbf{CAT} is unique in its focus on conditional association testing, we also provide results from the following marginal testing methods: a basic Mann-Whitney test, bias-corrected ANCOM  \citep[ANCOM-BC,][]{LinPedd20}, and DESeq2 \citep{LoveHube14}. Both the original and adjusted $p$-values for the DESeq2 method were computed. However, these methods have a different null hypothesis than  \textbf{CAT}, so cannot be considered as direct competitors.


	\subsection{Results}
 Here, we report findings from the \textbf{CAT} test paired with PERMANOVA using the weighted UniFrac metric for $\lambda=70$ for Porphyromonadaceae and $\lambda=10$ for Lachnospiraceae across 200 simulated datasets. 
 We chose to focus on these settings as all methods achieved high power to detect the spiked-in feature, but had differential results regarding the significance of its parent and child nodes.
 The results from \textbf{CAT} for other $\lambda$ values can be found in the Supplemental Material. 
Omitting sequences from the spiked-in feature or its parent node results in a sharp reduction in $R^2$ values for both synthetic data sets (Figure \ref{Sim} at left). The effect of omitting sequences from the child nodes is more nuanced; in the first data set, the child nodes of Porphyromonadaceae are responsible for explaining some portion of the $R^2$ value, while the child nodes of Lachnospiraceae may contribute less independent information. The percent of $p$-values less than 0.05  for the
\textbf{CAT} test along with the results from existing marginal tests are shown at right in Figure \ref{Sim}. 
In addition to the family being "spiked-in", i.e., the feature with an abundance difference constructed between the groups under the simulation design, we also offer the hypothesis testing results for the order above and the genera below.  The proposed \textbf{CAT} method can correctly reject the null hypothesis for the family directly manipulated and the order above most of the time.
In some cases, the $p$-values obtained from \textbf{CAT} are congruent with those obtained from marginal tests. However, for some hypotheses the conditional testing approach of  \textbf{CAT} is relatively more conservative than marginal tests. In particular, the genus \textit{Butyrivibrio}, a child of the spiked-in feature, is consistently found to be significant by ANCOM-BC and DESeq2, while the \textbf{CAT} results indicate that this feature is not responsible for driving the global association results.
This behavior of \textbf{CAT} reflects the nature of the conditional test; it will be less likely to reject the null hypothesis when other features have already explained the cross-group differences. In contrast, other tests estimate the marginal effect, which may overempasize the importance of lower-level taxa. 

	\section{Application to real data} \label{sec:application}
	We now discuss the use of \textbf{CAT} to analyze two real data sets examining the role of the microbiome in shaping melanoma patient outcomes. First, we consider the 
  data set from \cite{GopaSpen18}, which dealt with the association of the microbiome to immunotherapy response. In Section \ref{sec:spen}, to illustrate the use of  \textbf{CAT} for time-to-event outcomes, we apply \textbf{CAT} with MiRKAT-S to the study of \cite{SpenMcQu21}, which characterized the role of the microbiome in shaping progression free survival.

 \subsection{Binary response}\label{sec:deepak}

	In our first case study, we apply the \textbf{CAT} method to the melanoma dataset described in \cite{GopaSpen18}.
	In this data set, there are global differences in microbiome composition between patients that responded to immunotherapy vs.\ those that did not; 
	the $p$-value from the PERMANOVA test using weighted UniFrac for responder vs.\ nonresponder is less than 0.001.
 To identify specific differentially abundant features, 
	\cite{GopaSpen18} relied on LEfSe \citep{SegaIzar11}, the first step of which is to screen features with a Mann-Whitney test. We applied \textbf{CAT} to ascertain whether the hits they identified using LEfSe remained significant under a conditional test. 

	\begin{table}
	\begin{tabular}{l|l|r|r|r}\hline
		Level& Taxon& MW &$R^2$ & \textbf{CAT} \\
  		&& $p$-value&difference&  $p$-value\\
		\hline
    	Phylum &Bacteroidetes &$\mathbf{<0.01}$ &0.0502	&$0.072$\\
		Phylum &Firmicutes&$\mathbf{<0.01}$&0.0337	&$\mathbf{0.035}$\\
		Class &Bacteroidia&$\mathbf{<0.01}$	&0.0503&$0.072$\\
  	Class &Clostridia&$\mathbf{<0.01}$&0.0304	&$\mathbf{0.050}$\\
    	Class &Mollicutes&$\mathbf{0.01}$&0.0001	& $0.084$\\
		Order &Bacteroidales&$\mathbf{<0.01}$&0.0503	&$0.072$\\
		Order &Clostridiales&$\mathbf{<0.01}$&0.0303	&$\mathbf{0.050}$\\
  	Family &Micrococcaceae&$\mathbf{0.01}$&$<0.0001$	&$0.071$\\
		Family & Ruminococcaceae&$\mathbf{0.03}$ & 0.0365 &$\mathbf{<0.001}$\\
    	Genus &\textit{Faecalibacterium}&$\mathbf{0.01}$&0.0151	&$\mathbf{0.005}$\\
        Genus &\textit{Gardnerella}&$\mathbf{0.03}$&	$<0.0001$&$0.879$ \\
        Genus &\textit{Peptoniphilus} &0.12&$<0.0001$&$0.148$ \\
        Genus &\textit{Phascolarctobacterium}&$\mathbf{0.01}$&0.0010&$\mathbf{0.029}$\\
        Genus &\textit{Rothia}	&$\mathbf{0.01}$ & $<0.0001$&$0.071$\\
        Genus &\textit{Ruminococcus}&$\mathbf{0.03}$ & 0.0096	&$\mathbf{<0.001}$\\
        Species &\textit{B.\ stercoris}&$\mathbf{0.03}$	&$<0.0001$&$0.866$\\
		Species &\textit{F.\ prausnitzii}&$\mathbf{0.01}$	&0.0151&$\mathbf{0.005}$\\
		Species &\textit{M.\ hungatei}&0.18	&$<0.0001$&$\mathbf{0.038}$\\
		Species &\textit{R.\ bromii}&0.08	&0.0043&$\mathbf{0.002}$\\\hline
	\end{tabular} \vskip.2cm
\caption{Level in the taxonomic tree, taxon, Mann-Whitney $p$-value, $R^2$ difference before and after removing candidate taxon, and $p$-value from \textbf{CAT} when applied to features identified by LEfSe in \cite{GopaSpen18}.}
\label{DeepakTable}
 \end{table}

\begin{figure}
    \centering
     \includegraphics[width=\textwidth]{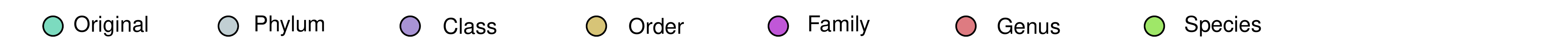}\\
    \hskip-2cm
\vskip-0.4cm     \hskip-0.5cm
\includegraphics[width=\textwidth]{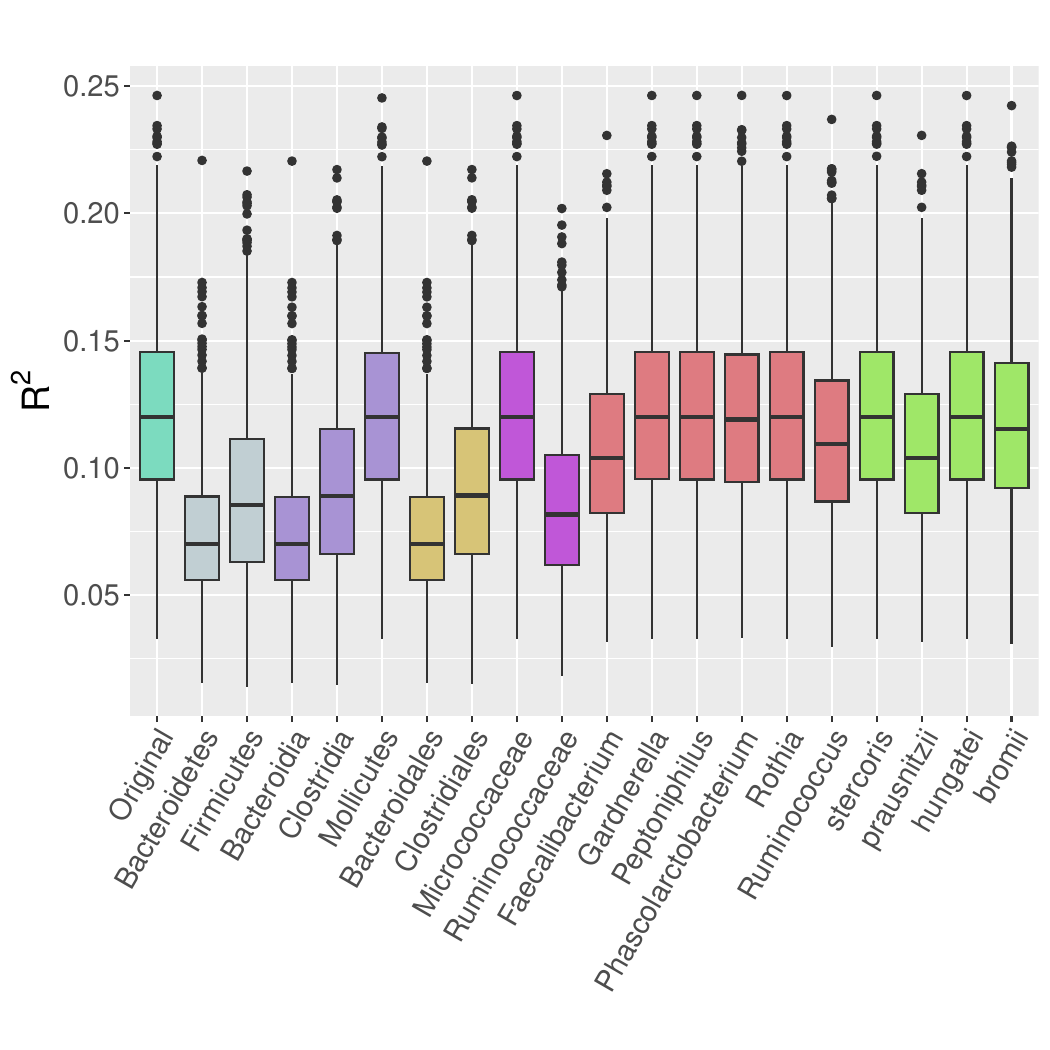}
   \caption{Application of \textbf{CAT} to \cite{GopaSpen18}'s dataset described in Section \ref{sec:deepak}}
    \label{Deepak}
\end{figure}

Table \ref{DeepakTable} shows the Mann-Whitney $p$-value as well as the $R^2$ difference and $p$-value obtained using \textbf{CAT}; significant hits from  \textbf{CAT} include the Phylum Firmicutes, its subordinate class Clostridia, the order Clostridiales under Clostridia, the family Ruminococcaceae within Clostridiales, and the genus \textit{Ruminococcus} that comes under Ruminococcaceae. Additionally, the species \textit{R.\ bromii} beneath \textit{Ruminococcus} is deemed significant. Within the same family, Ruminococcaceae, the genus Faecalibacterium, and the species underneath, prausnitzii, are also significant. 
However, some hits that were found to be significant using LEfSe, including the genus \textit{Gardnerella} and the species \textit{B. stercoris} lose significance; this suggests that these microbiome features might not be good candidates for a microbiome intervention. The difference in $R^2$ values across bootstrap iterations is illustrated in Figure \ref{Deepak}. This figure further suggests that intervening on higher-level taxa, such as Ruminococcaceae, might be expected to exert a larger influence than designing an intervention focused on species-level hits. Overall, \textbf{CAT} provides many results that are congruent with the original paper, yet provides novel insights into the conditional association.

\subsection{Survival outcomes}\label{sec:spen}
To demonstrate the application of \textbf{CAT} with the MiRKAT-S method for survival outcomes, we employed it to analyze the dataset from \cite{SpenMcQu21}. This dataset included 163 subjects undergoing systemic therapy for melanoma that were profiled using 16S rRNA sequencing and followed for progression free survival. Among these subjects, 86 progression events were observed, with a median PFS of 1.8 years.
The microbiome profiling data included 3306 ASVs, corresponding to 346 unique taxa at the genus level or higher. 
In our case study, we focused on taxa found to be associated with treatment response by \cite{SpenMcQu21}, including the phylum Firmicutes, class Clostridia, order Oscillospirales, family Ruminococcaceae, and genera \textit{Faecalibacterium} and \textit{Ruminococcus}. We also tested the genera \textit{Bifidobacterium} and \textit{Lactobacillus}, as these are popular in commercially available supplements and were tested as probiotic interventions as part of a pre-clinical experiment in the same study \citep{SpenMcQu21}.
 To apply \textbf{CAT}, we used Bray-Curtis and Jaccard distances with the MiRKAT-S method. In addition, we ran the univariate Cox model for each candidate feature for comparison.

\begin{table}
	\begin{tabular}{l|l|r|r|r}\hline
			Level& Taxon&Cox &$R^2$ & \textbf{CAT} \\
   			Level& Taxon&$p$-value&difference& $p$-value\\\hline
   Phylum&	Firmicutes&	$0.54$	&$0.0011$&	$\mathbf{0.0018}$\\
   Class&	Clostridia&	$0.46$&	$0.0010$	&$\mathbf{0.0033}$\\
   Order&	Oscillospirales&	$0.18$&	$0.0007$&	$\mathbf{0.0364}$\\
Family&	Ruminococcaceae&	$\mathbf{0.04}$&$	0.0006$&	$0.1526$\\
Genus&	\textit{Faecalibacterium}&	$\mathbf{0.02}$&	$0.0005$&	$0.3135$\\
Genus&	\textit{Ruminococcus}&	$0.31$&$	<0.0001$&$	0.1526$\\
Genus&	\textit{Bifidobacterium}	&$\mathbf{0.02}$&$	<0.0001$&	$0.3422$\\
Genus&	\textit{Lactobacillus}&	$0.58$&$	<0.0001$&$	0.2979$\\\hline
		\end{tabular} \vskip.2cm
	\caption{Level in the taxonomic tree, taxon, univariate Cox model $p$-value, $R^2$ difference before and after removing candidate taxon, and $p$-value from \textbf{CAT} when applied to features of interest from \cite{SpenMcQu21}.}
	\label{Spencer}
\end{table}

Table \ref{Spencer} displays the outcomes of our \textbf{CAT} method. Remarkably, \textbf{CAT} identifies features closer to the leaf nodes of the taxonomic tree as non-significant, as such features may not add explanatory information for the variability of the outcome. 
However, \textbf{CAT} finds statistical significance for higher-level taxa, encompassing order Oscillospirales, the class Clostridia above it, and the phylum Firmicutes above the class. In contrast, when using Cox models for marginal tests, the finer-resolution taxonomic units are deemed significant.

	\section{Concluding remarks}\label{sec:discussion}
To date, most existing methods for microbiome association testing have focused on either global or marginal testing. Here, we adopt a conditional testing framework, proposing the \textbf{CAT} method as a conditional association test using a leave-out approach. The leave-out idea is one of the most fundamental ideas in statistical testing,  whose applications range from likelihood ratio testing to type III sum-of-squares analysis. \textbf{CAT} combines this classic idea with the flexibility of using various metrics and testing approaches designed for microbiome data. It is worth mentioning that though microbiome data motivate us to propose \textbf{CAT}, the method is widely applicable to a wide range of situations where non-Euclidean pairwise distances are used. In this paper, we only illustrate how to test the conditional association for one taxon; testing the effect of several taxa as a unit is also possible by changing leave-one-out to leave-multiple-out in Step 3 of the procedure.

Although our simulation results show that \textbf{CAT} may identify fewer features than marginal tests, particularly at lower levels in the tree, we do not directly address multiple testing in this paper. Commonly used multiplicity adjustments, such as the Bonferroni procedure or Benjamini-Hochberg procedure can be applied to $p$-values generated by \textbf{CAT}. However, the null hypotheses under testing in microbiome data sets are not independent. False discovery rate control in correlated conditional tests, particularly in the presence of a phylogenetic tree structure, is an area that we plan to address in the  future.

The results from the \textbf{CAT} method have clear real-world relevance. There is a growing effort to develop interventions aimed at reshaping the microbiome by administering "rationally designed" mixtures of bacterial strains \citep{Van2021} which have been selected to confer potential benefit to the patient. Our method can identify features that are potentially influential conditioning on the existence of other bacteria. This will help clinicians identify intervention targets more efficiently.


	\section*{Funding}
	KAD is partially supported by NIH P30CA016672, SPORE P50CA140388, CCTS TR000371 and CPRIT
RP160693. CBP is partially supported by NIH R01 HL158796 and NIH/NCI CCSG P30CA016672. RRJ is partially supported by NIH R01 HL124112 and NIH R01 HL158796. YS is partially supported by NSF DMS 2310955.
	
	\bibliographystyle{natbib}

\end{document}


\begin{center}
	\textbf{\Large Supplement to ``CAT: A conditional association test for \\ \vskip.2cm microbiome data using a leave-out approach"} \\ \vskip.2cm 
	\textit{Yushu Shi, Liangliang Zhang, Kim-Anh Do, Robert R.\ Jenq, and \\ Christine B. Peterson} \\ \vskip.4cm
\end{center}
\vskip.5cm
\section*{Additional simulation results}
Here, we provide additional simulation results with $\lambda=5,10,30,50$, and $70$. For a detailed description of the simulation design and methods compared, please see Section 3 of the main manuscript.

\newpage
\subsection*{$\lambda = 5$}
	\begin{figure}[h!]
 \centering
		\begin{subfigure}{\linewidth}
			\caption{Family Porphyromonadaceae and related taxa}
			\vskip-.16cm 
   \includegraphics*[width=0.46\textwidth,trim={0 0 0 1cm},clip]{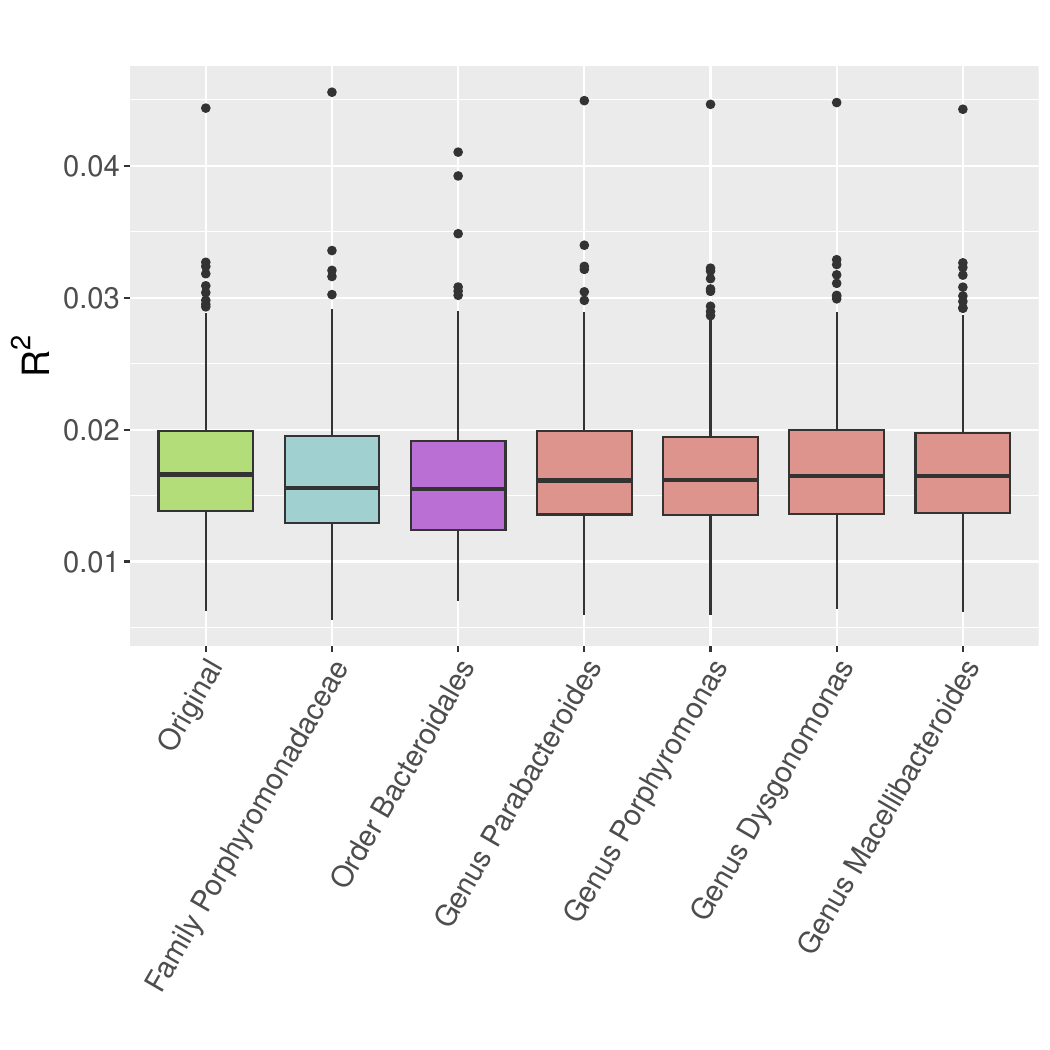}
			\includegraphics*[width=0.48\textwidth,trim={0 0 0 1cm},clip]{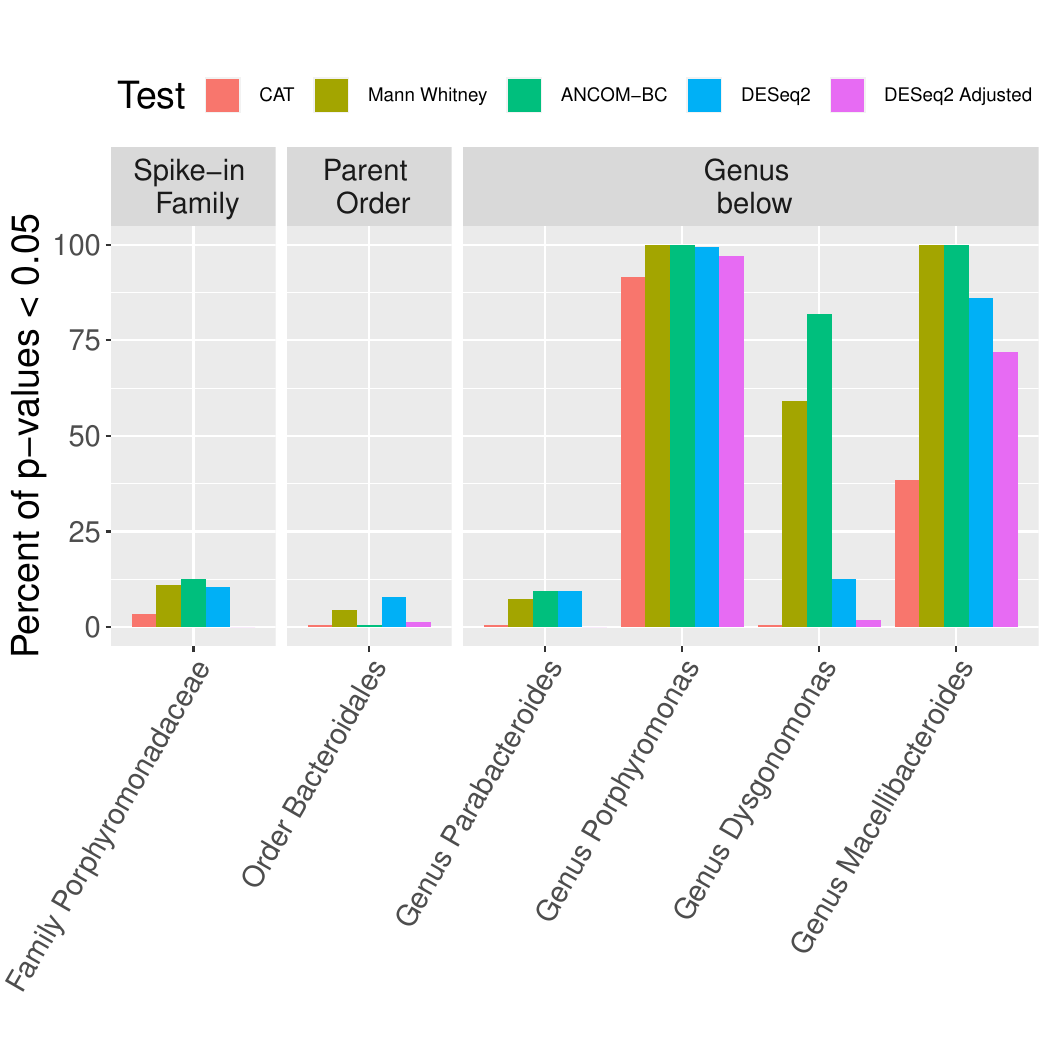}
		\end{subfigure}
		
		\begin{subfigure}{\linewidth}
			\vskip-.4cm  \caption{Family Lachnospiraceae and related taxa}
				\vskip-.16cm 
      \includegraphics*[width=0.46\textwidth,trim={0 0 0 1cm},clip]{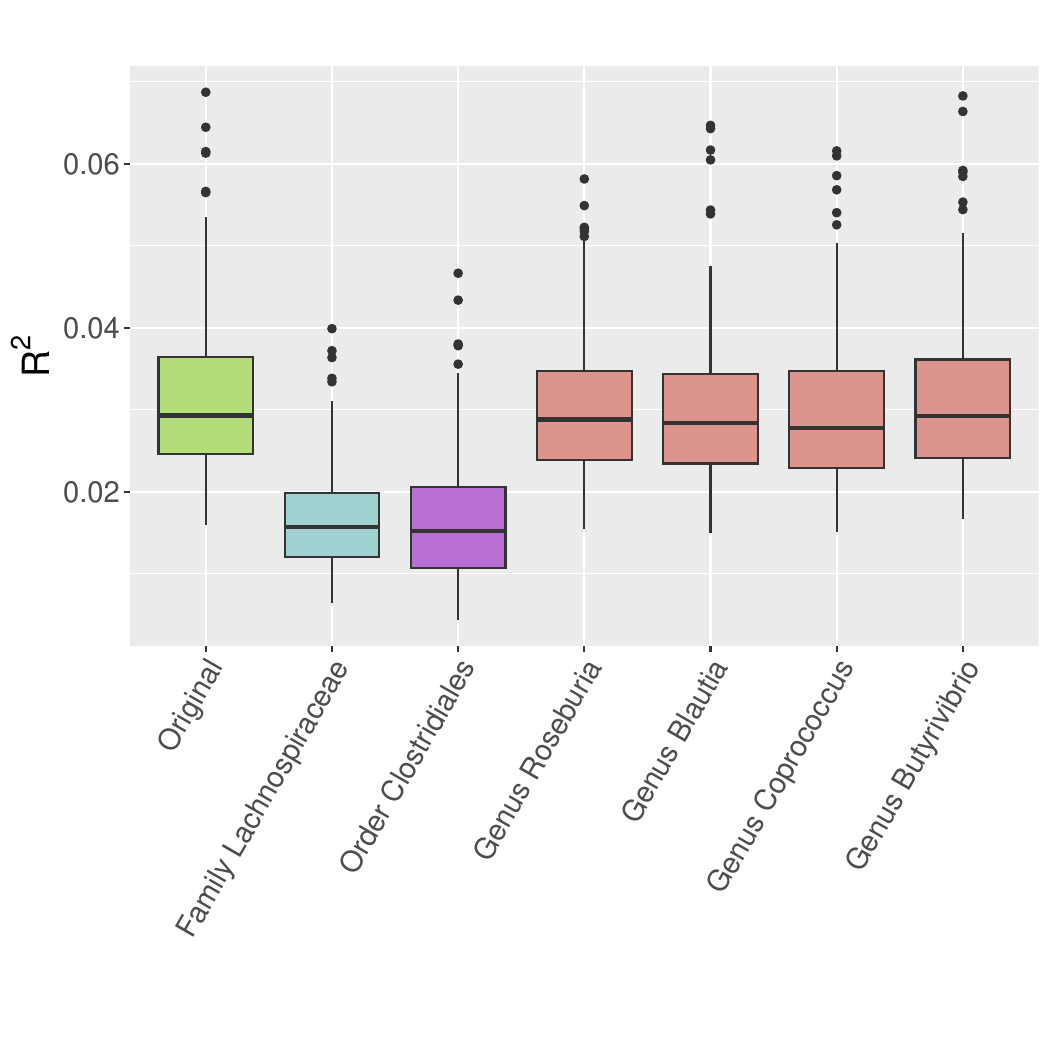}
			\includegraphics*[width=0.48\textwidth,trim={0 0 0 1cm},clip]{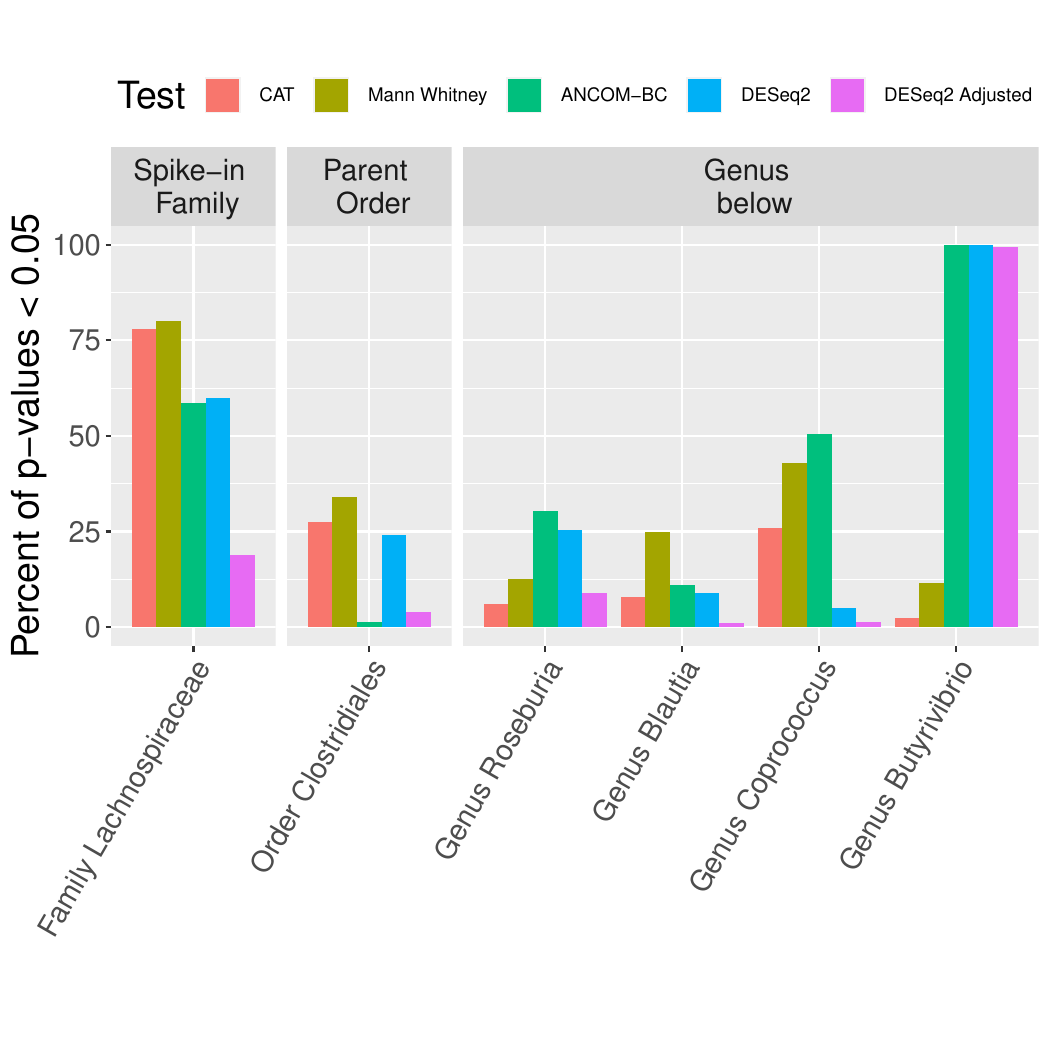}
		\end{subfigure}
		\caption{Boxplots of the $R^2$ values (left) and 
  barplots of the percentage of $p$-values less than 0.05 from \textbf{CAT}, Mann-Whitney, ANCOM-BC, and DeSeq2 (right) over 200 simulated datasets with $\lambda = 5$. The first panel in each subplot in the right column represents results from the manipulated family, followed by the order above and the child genera below.}
		\label{Sim}
	\end{figure}

\newpage
\subsection*{$\lambda = 10$}
	\begin{figure}[h!]
 \centering
		\begin{subfigure}{\linewidth}
			\caption{Family Porphyromonadaceae and related taxa}
			\vskip-.16cm 
   \includegraphics*[width=0.46\textwidth,trim={0 0 0 1cm},clip]{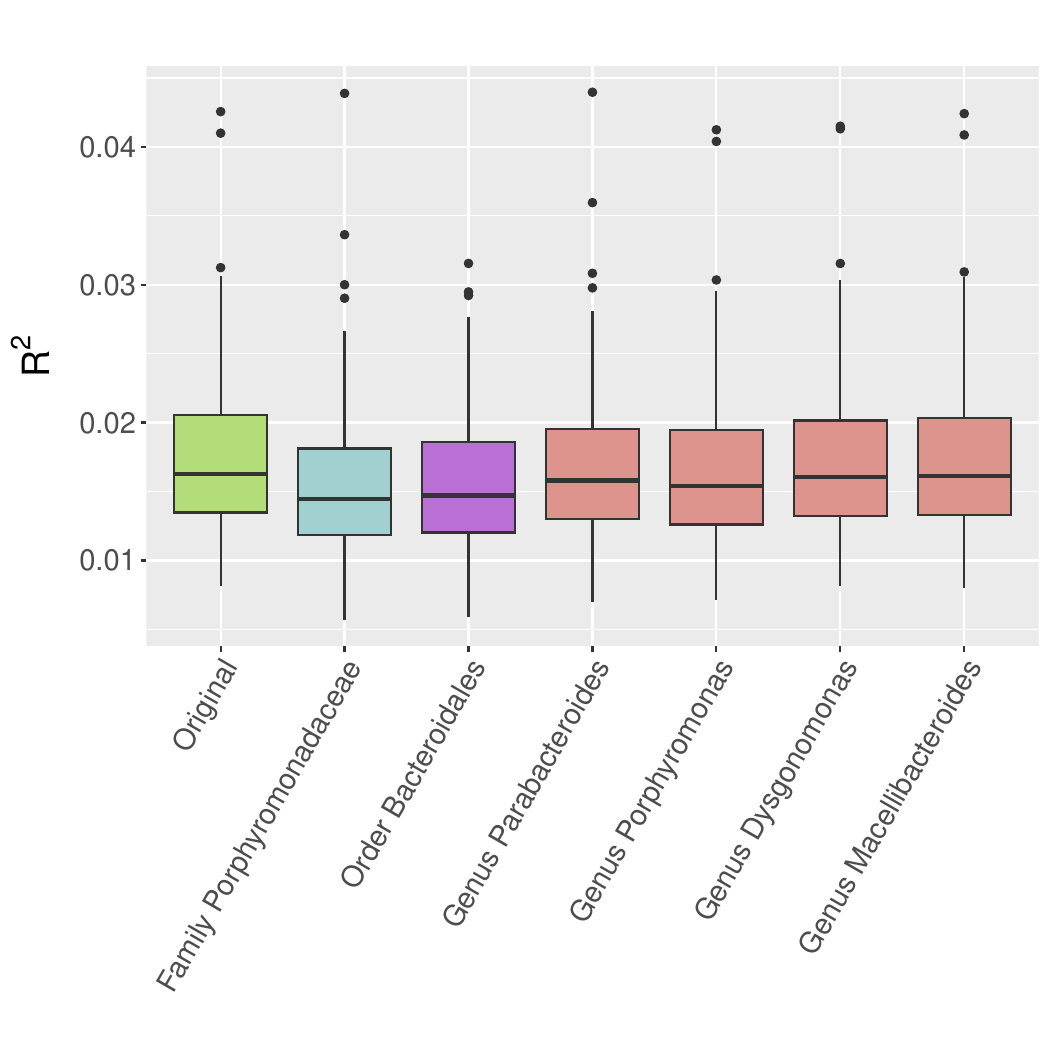}
			\includegraphics*[width=0.48\textwidth,trim={0 0 0 1cm},clip]{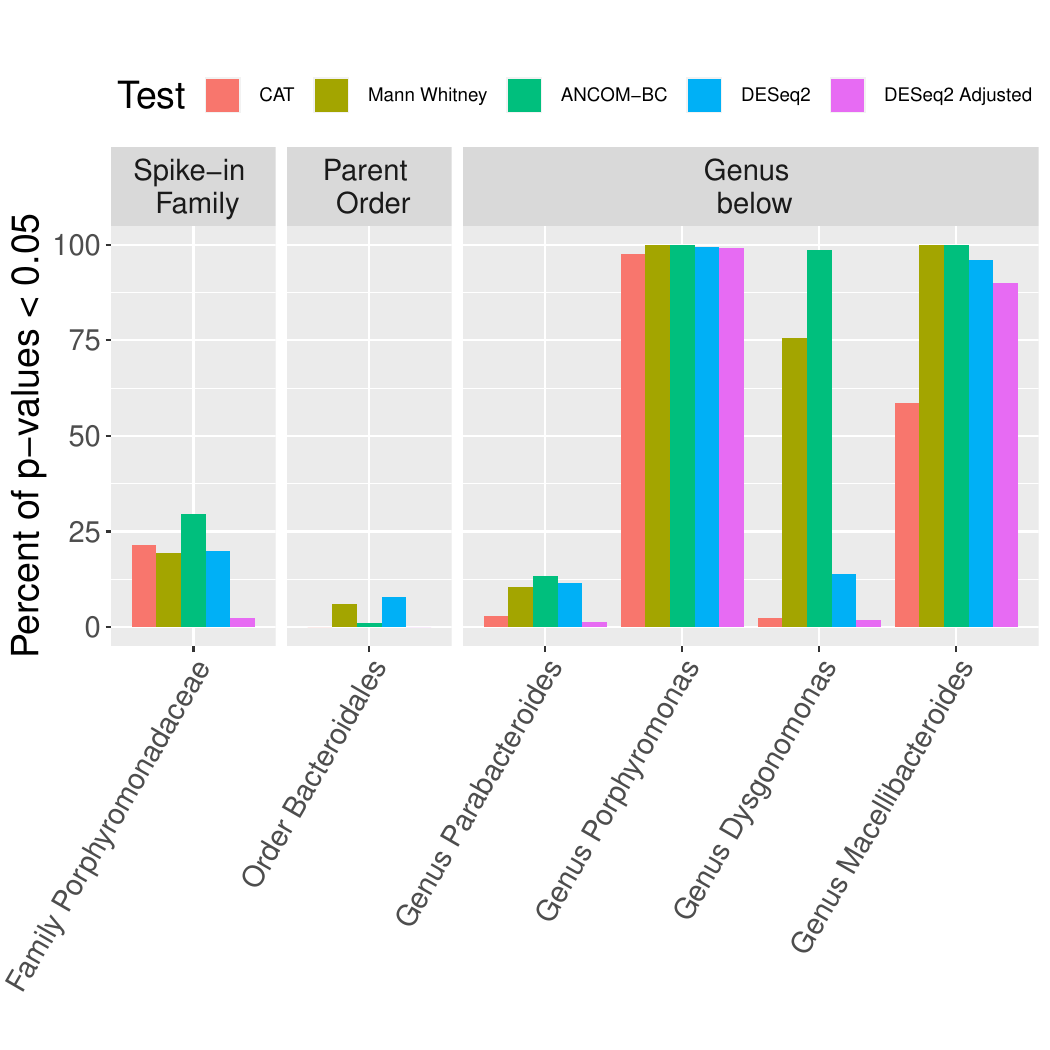}
		\end{subfigure}
		
		\begin{subfigure}{\linewidth}
			\vskip-.4cm  \caption{Family Lachnospiraceae and related taxa}
				\vskip-.16cm 
      \includegraphics*[width=0.46\textwidth,trim={0 0 0 1cm},clip]{plots/boxPlotCBPLach10BarR2}
			\includegraphics*[width=0.48\textwidth,trim={0 0 0 1cm},clip]{plots/CBPLach10BarR2}
		\end{subfigure}
		\caption{Boxplots of the $R^2$ values (left) and 
  barplots of the percentage of $p$-values less than 0.05 from \textbf{CAT}, Mann-Whitney, ANCOM-BC, and DeSeq2 (right) over 200 simulated datasets with $\lambda = 10$. The first panel in each subplot in the right column represents results from the manipulated family, followed by the order above and the child genera below.}
		\label{Sim}
	\end{figure}

 \newpage
\subsection*{$\lambda = 30$}
	\begin{figure}[h!]
 \centering
		\begin{subfigure}{\linewidth}
			\caption{Family Porphyromonadaceae and related taxa}
			\vskip-.16cm 
   \includegraphics*[width=0.46\textwidth,trim={0 0 0 1cm},clip]{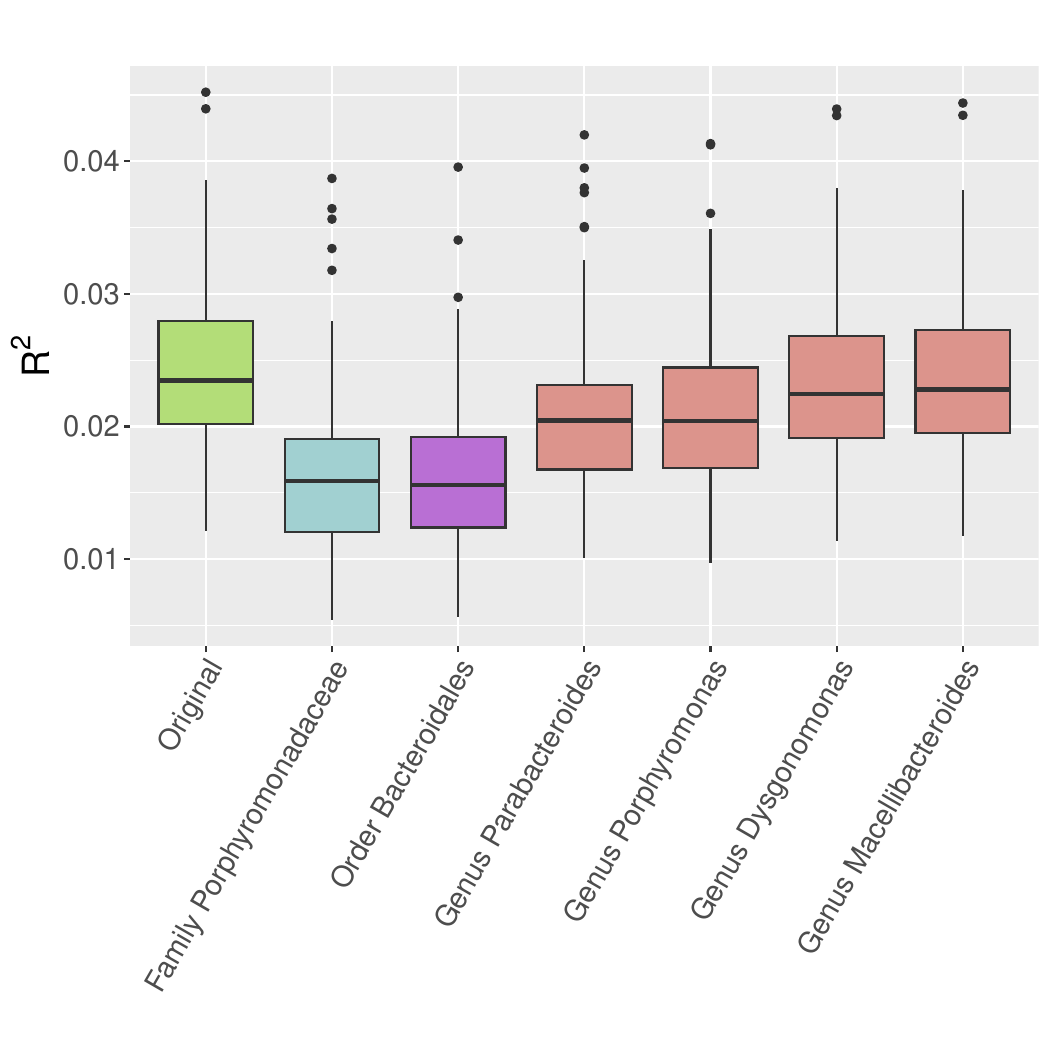}
			\includegraphics*[width=0.48\textwidth,trim={0 0 0 1cm},clip]{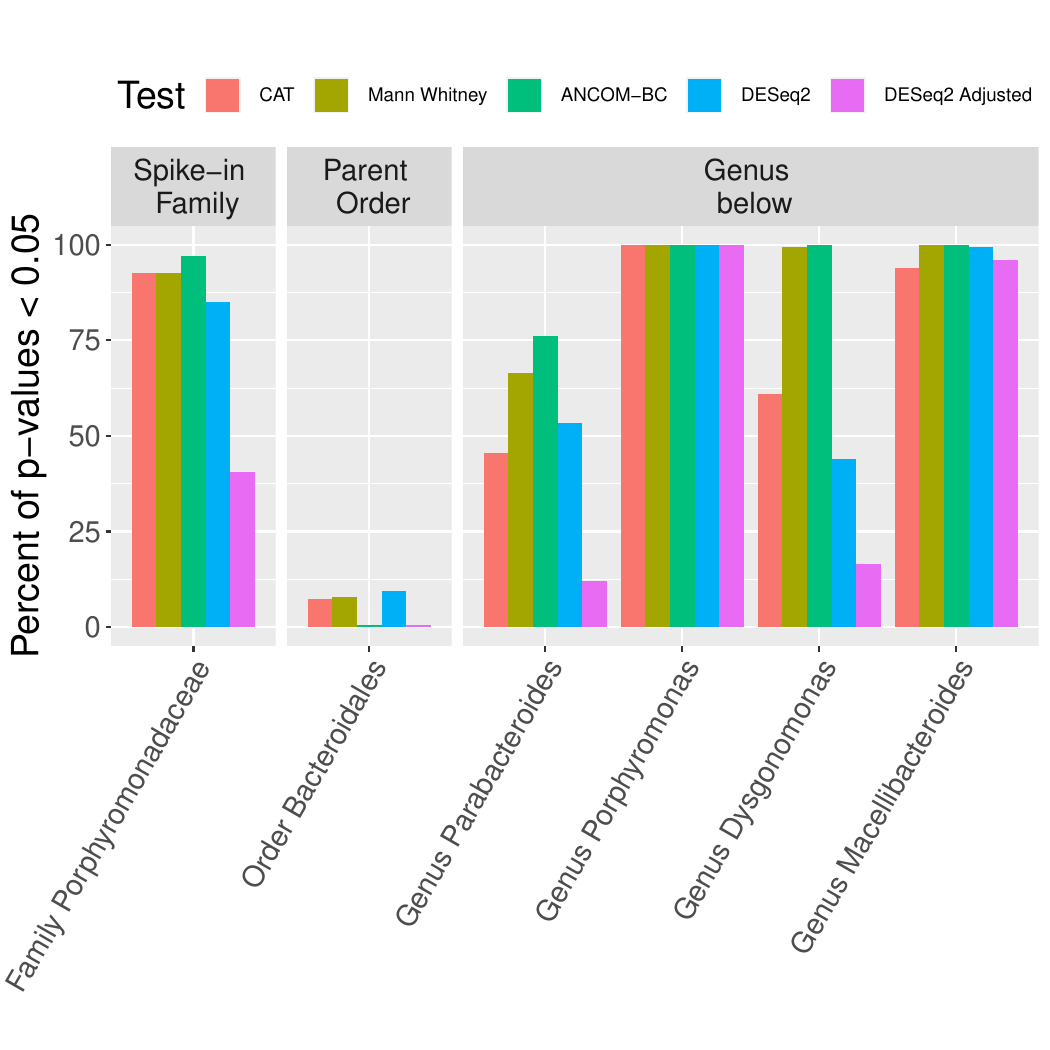}
		\end{subfigure}
		
		\begin{subfigure}{\linewidth}
			\vskip-.4cm  \caption{Family Lachnospiraceae and related taxa}
				\vskip-.16cm 
      \includegraphics*[width=0.46\textwidth,trim={0 0 0 1cm},clip]{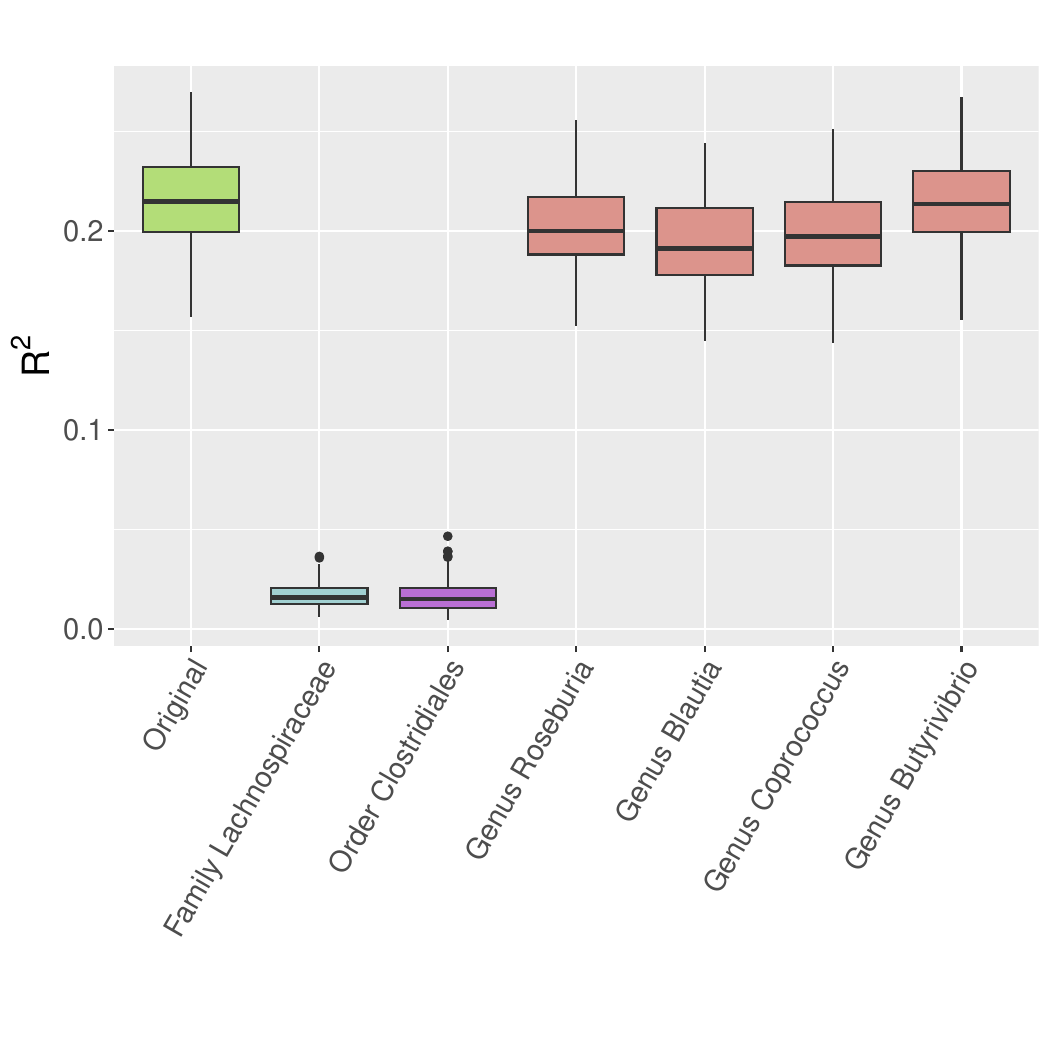}
			\includegraphics*[width=0.48\textwidth,trim={0 0 0 1cm},clip]{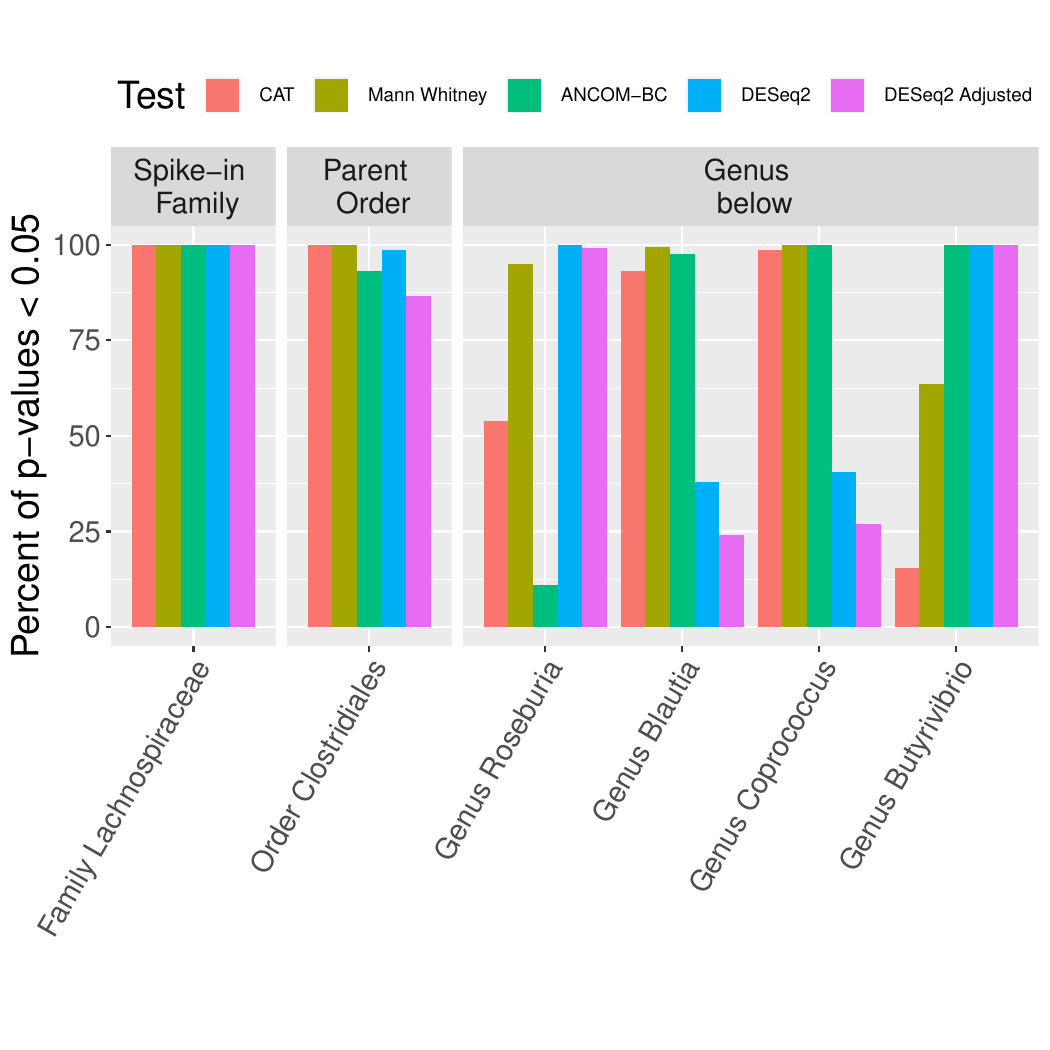}
		\end{subfigure}
		\caption{Boxplots of the $R^2$ values (left) and 
  barplots of the percentage of $p$-values less than 0.05 from \textbf{CAT}, Mann-Whitney, ANCOM-BC, and DeSeq2 (right) over 200 simulated datasets with $\lambda = 30$. The first panel in each subplot in the right column represents results from the manipulated family, followed by the order above and the child genera below.}
		\label{Sim}
	\end{figure}

  \newpage
\subsection*{$\lambda = 50$}
	\begin{figure}[h!]
 \centering
		\begin{subfigure}{\linewidth}
			\caption{Family Porphyromonadaceae and related taxa}
			\vskip-.16cm 
   \includegraphics*[width=0.46\textwidth,trim={0 0 0 1cm},clip]{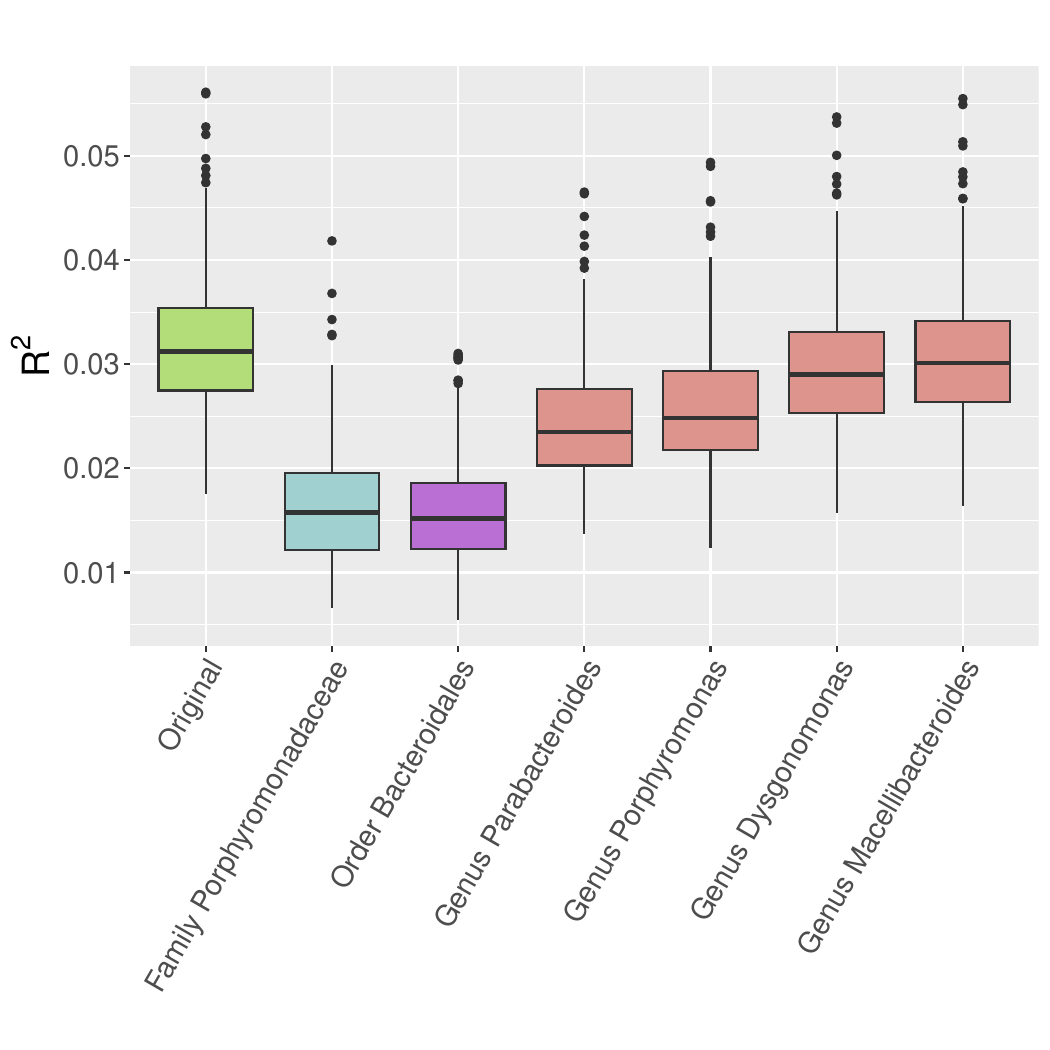}
			\includegraphics*[width=0.48\textwidth,trim={0 0 0 1cm},clip]{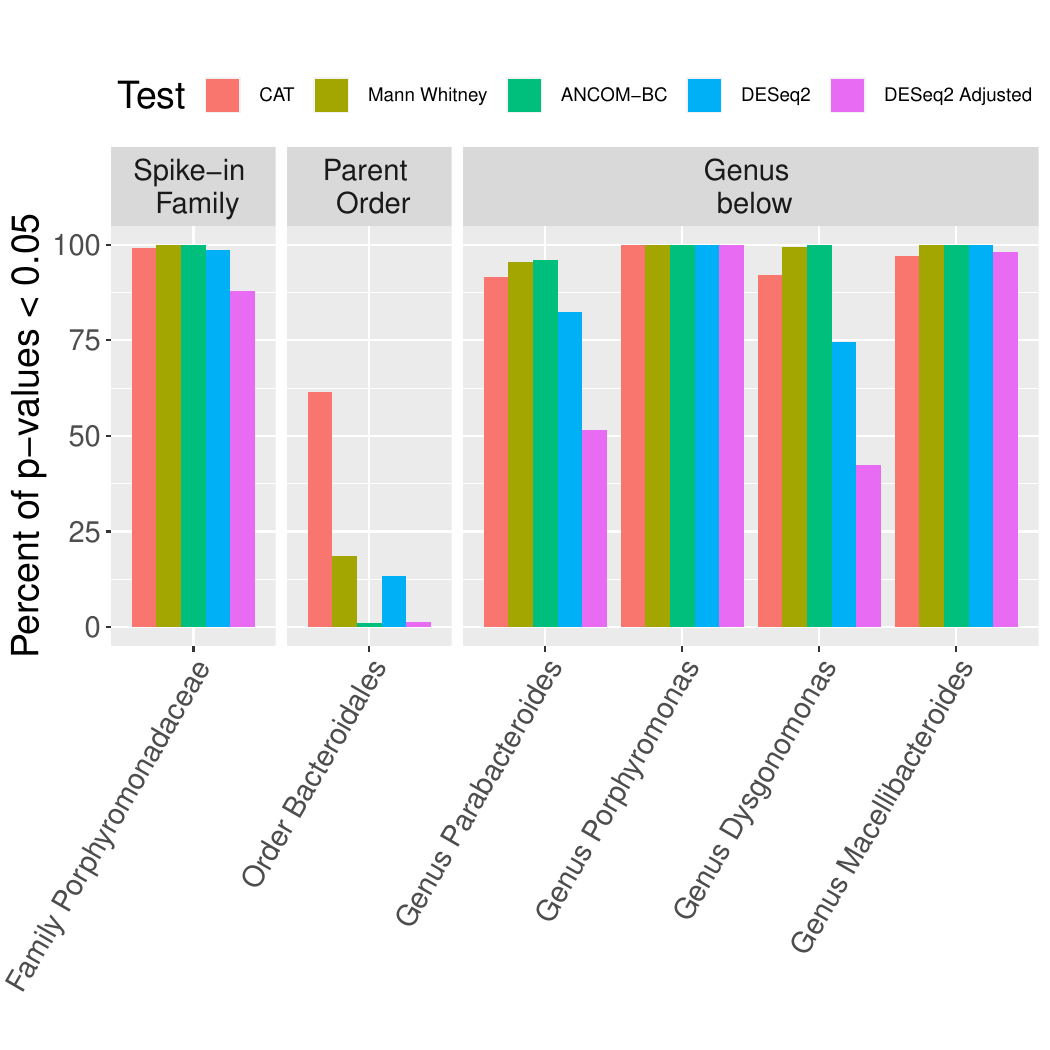}
		\end{subfigure}
		
		\begin{subfigure}{\linewidth}
			\vskip-.4cm  \caption{Family Lachnospiraceae and related taxa}
				\vskip-.16cm 
      \includegraphics*[width=0.46\textwidth,trim={0 0 0 1cm},clip]{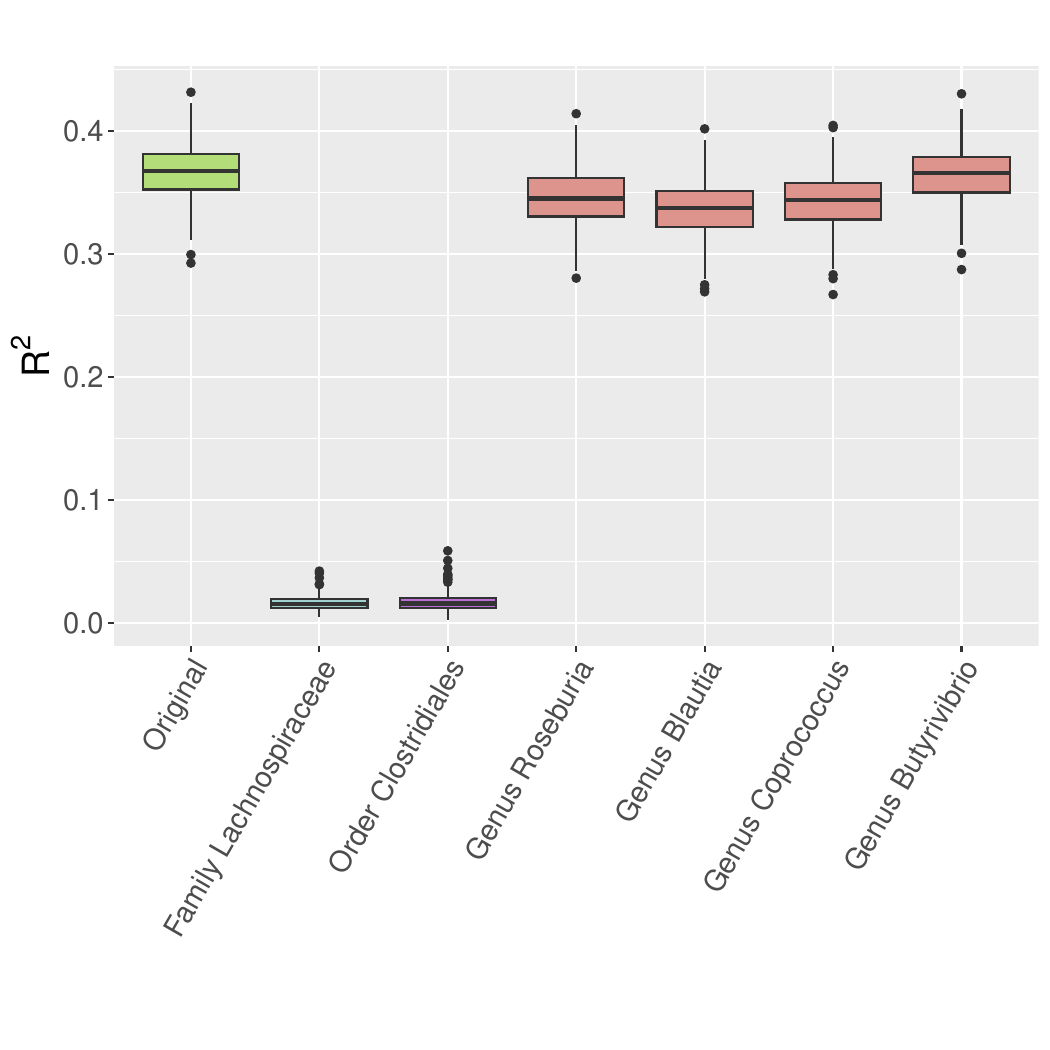}
			\includegraphics*[width=0.48\textwidth,trim={0 0 0 1cm},clip]{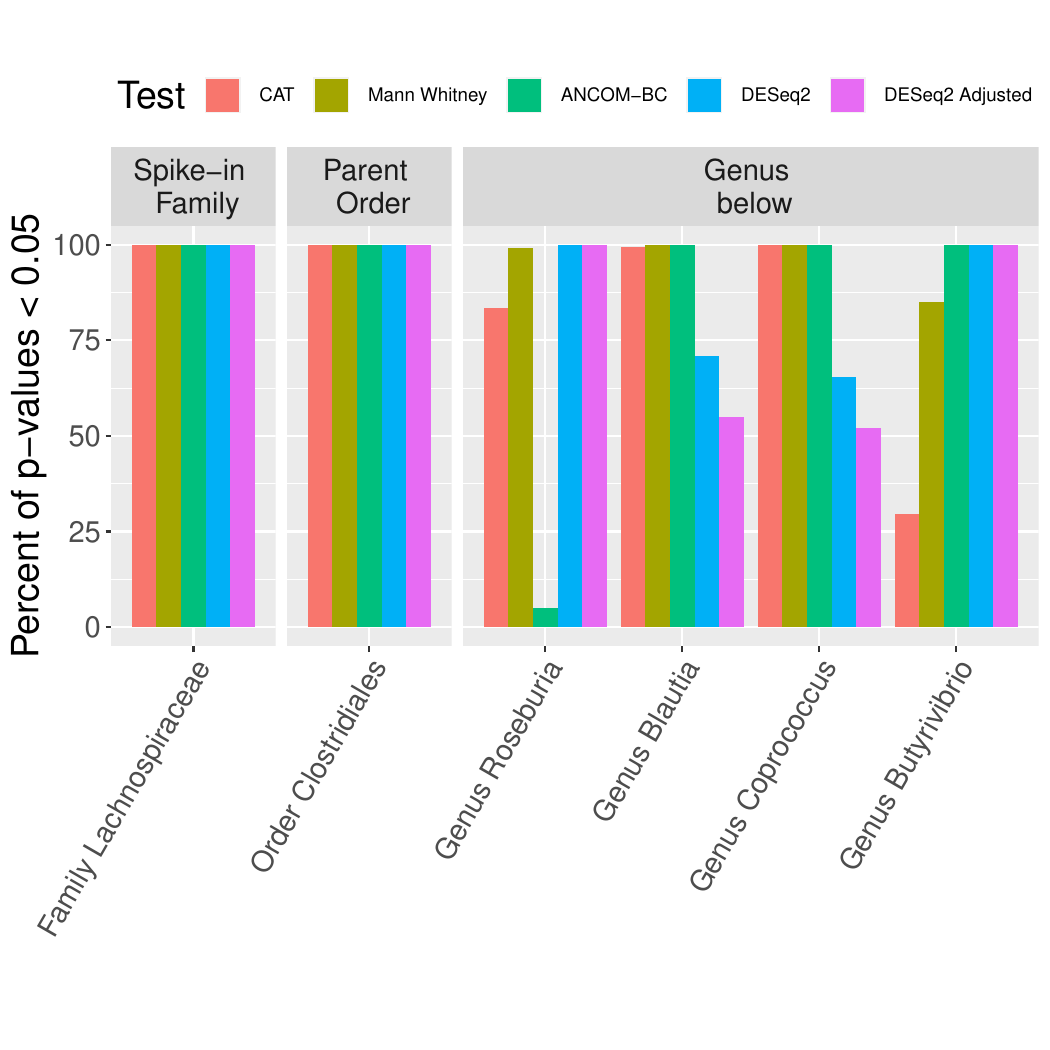}
		\end{subfigure}
		\caption{Boxplots of the $R^2$ values (left) and 
  barplots of the percentage of $p$-values less than 0.05 from \textbf{CAT}, Mann-Whitney, ANCOM-BC, and DeSeq2 (right) over 200 simulated datasets with $\lambda = 50$. The first panel in each subplot in the right column represents results from the manipulated family, followed by the order above and the child genera below.}
		\label{Sim}
	\end{figure}

   \newpage
\subsection*{$\lambda = 70$}
	\begin{figure}[h!]
 \centering
		\begin{subfigure}{\linewidth}
			\caption{Family Porphyromonadaceae and related taxa}
			\vskip-.16cm 
   \includegraphics*[width=0.46\textwidth,trim={0 0 0 1cm},clip]{plots/boxPlotCBPPorphy70BarR2}
			\includegraphics*[width=0.48\textwidth,trim={0 0 0 1cm},clip]{plots/CBPPorphy70BarR2}
		\end{subfigure}
		
		\begin{subfigure}{\linewidth}
			\vskip-.4cm  \caption{Family Lachnospiraceae and related taxa}
				\vskip-.16cm 
      \includegraphics*[width=0.46\textwidth,trim={0 0 0 1cm},clip]{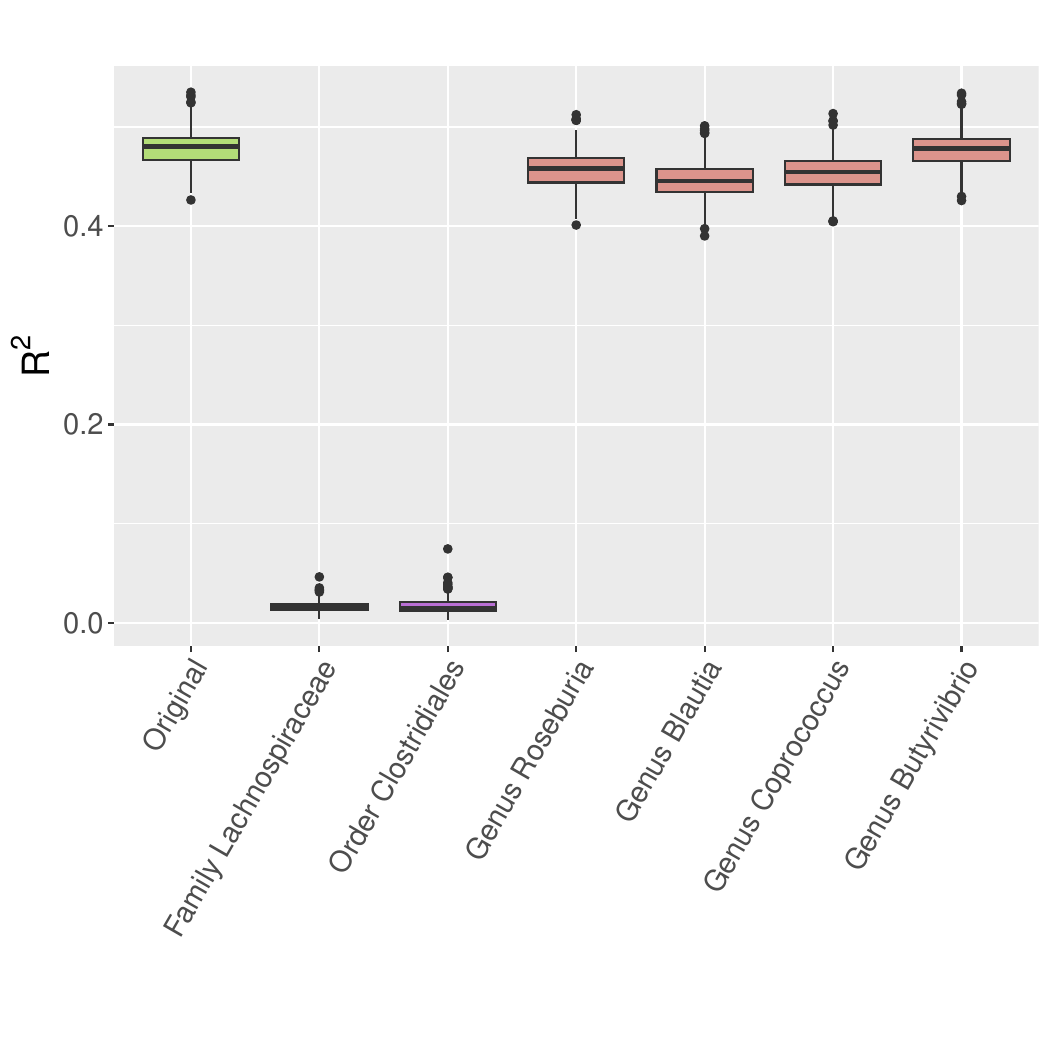}
			\includegraphics*[width=0.48\textwidth,trim={0 0 0 1cm},clip]{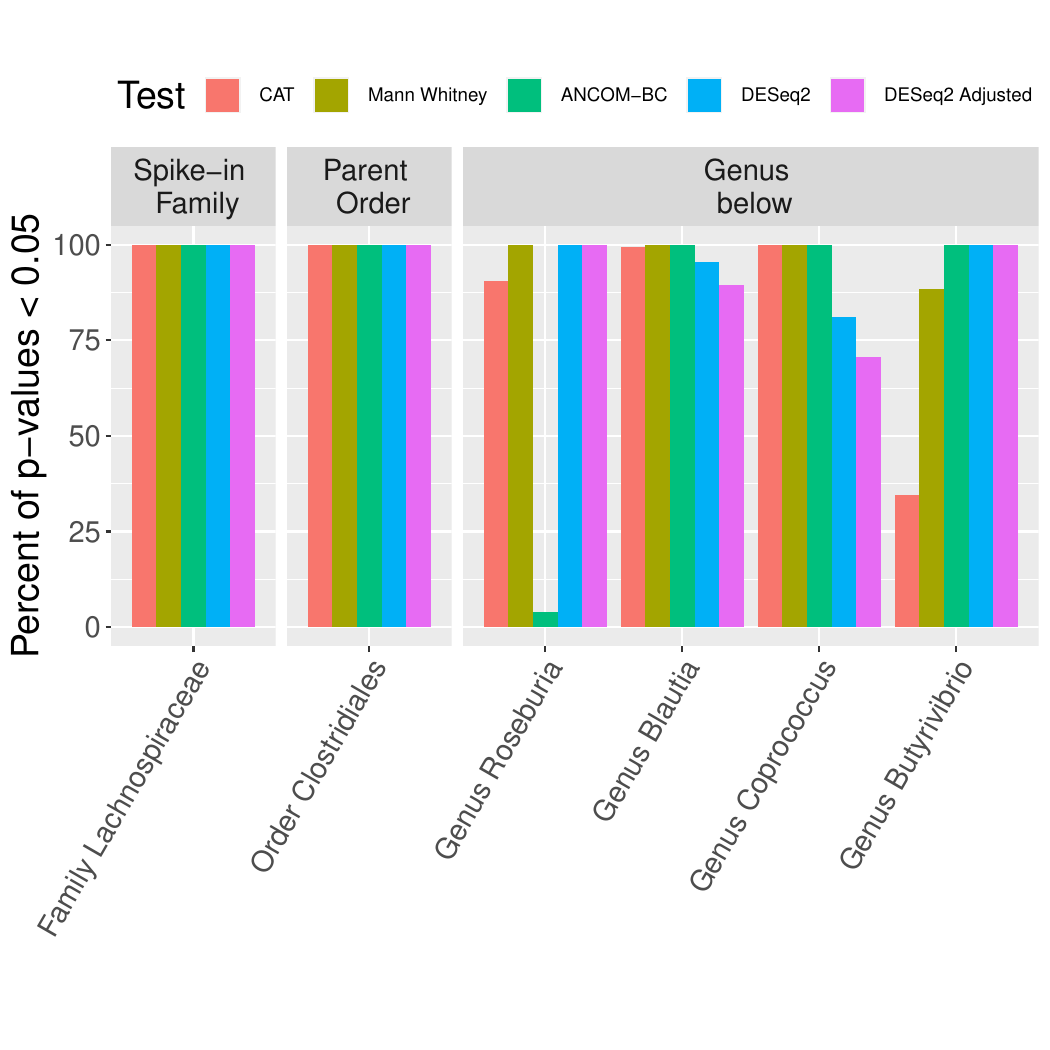}
		\end{subfigure}
		\caption{Boxplots of the $R^2$ values (left) and 
  barplots of the percentage of $p$-values less than 0.05 from \textbf{CAT}, Mann-Whitney, ANCOM-BC, and DeSeq2 (right) over 200 simulated datasets with $\lambda = 70$. The first panel in each subplot in the right column represents results from the manipulated family, followed by the order above and the child genera below.}
		\label{Sim}
	\end{figure}



